\documentclass[preprint,12pt]{elsarticle}

\usepackage{amssymb}

\usepackage{lineno}
\usepackage{graphicx}
\usepackage{xcolor}
\usepackage{color}
\usepackage{amsmath}
\journal{Nuclear Instrumentation and Methods in Physics Research A}

\begin{document}

\begin{frontmatter}

\title{Hyperon signatures in the PANDA experiment at FAIR}

\renewcommand{\theaffn}{\arabic{affn}}

\author[]{The PANDA Collaboration}

%
%
\author[1]{Viktor Abazov}
\author[2]{Victor Abramov}
\author[3,4]{Patrick Achenbach}
\author[5]{Orestis Afedulidis}
\author[5]{Lennart Ahrens}
\author[6]{Adeel Akram}
\author[7,8]{Mohammad Al-Turany}
\author[5]{Malte Albrecht}
\author[1]{Gennady Alexeev}
% ...
\affiliation[1]{organization={Joint Institute for Nuclear Research},
city={Dubna},
country={Russia}}
\affiliation[2]{organization={A.A. Logunov Institute for High Energy Physics of the National Research Centre “Kurchatov Institute”},
city={Protvino},
country={Russia}}
\affiliation[3]{organization={Johannes Gutenberg-Universität Institut für Kernphysik},
city={Mainz},
country={Germany}}
\affiliation[4]{organization={Helmholtz-Institut Mainz},
city={Mainz},
country={Germany}}
\affiliation[5]{organization={Ruhr-Universität Bochum Institut für Experimentalphysik I},
city={Bochum},
country={Germany}}
\affiliation[6]{organization={Uppsala Universitet Institutionen för fysik och astronomi},
city={Uppsala},
country={Sweden}}
\affiliation[7]{organization={GSI Helmholtzzentrum für Schwerionenforschung GmbH},
city={Darmstadt},
country={Germany}}
\affiliation[8]{organization={European Organization for Nuclear Research (CERN)},
city={Geneva},
country={Switzerland}}
% ...

\begin{abstract}
We present a detailed simulation study of the signatures from the sequential decays of the triple-strange $\bar{p}p\rightarrow\bar{\Omega}^+\Omega^- \to K^+\bar{\Lambda}K^-\Lambda \to K^+\bar{p}\pi^+K^-p\pi^-$ process in the PANDA central tracking system with focus on hit patterns and precise time measurement. We present a systematic approach for studying physics channels at the detector level and develop input criteria for tracking algorithms and trigger lines. Finally, we study the beam momentum dependence on the reconstruction efficiency for the PANDA detector.
\end{abstract}




\end{frontmatter}


\section{Introduction}
\label{sec:intro}

\noindent Strange hyperons, \textit{i.e.} baryons containing at least one strange quark, have been proven to be powerful diagnostic tools for the strong interaction and fundamental symmetries \cite{bes3prl,bes3nature,clashyp,hadeshyp}. In the future PANDA (antiProton ANnihilation at DArmstadt) experiment at FAIR (Facility for Antiproton and Ion Research) \cite{panda} hyperon-antihyperon pairs will be produced in $\bar{p}p$ annihilations at $\bar{p}$ beam momenta ranging from 1.5 GeV/\textit{c} to 15.0 GeV/\textit{c}. The experiment has the potential to take the field of hyperon physics to a new level, as demonstrated for single- and double-strange hyperons in Refs. \cite{pandahyp,phaseone,pandaspec,pandahypn}. The broad kinematic range of the HESR accelerator and the versatility of the PANDA detector also enables studies of the triple-strange $\Omega^-$ hyperon. The $\Omega^-$ hyperon played a crucial role in the development of modern hadron physics: its experimental discovery in 1964 \cite{Omegafirst} confirmed its existence as predicted by the Eight-fold Way \cite{eightfold}. Large-scale studies of the $\Omega^-$ hyperon and its antiparticle, the $\bar{\Omega}^+$, offers unique opportunities to study strong interaction dynamics \cite{kaidalov} and fundamental symmetries \cite{perotti1,perotti2}. Furthermore, hyperon-antihyperon pair studies close to the kinematic threshold, where the produced pair is in a relative s-wave, provide a probe of the interaction potential. This, in turn, is an important component in the understanding of the hyperon internal structure \cite{haidenbauer} as well as the role of strangeness in macroscopic objects such as neutron stars \cite{neutronstar}. 

For hyperons decaying into a baryon and a pion (such as the $\Lambda$, the $\Sigma$ isotriplet states and the $\Xi$ isodoublet states), the major part of the excess energy is generally transferred to the baryon due to momentum conservation. As a consequence, near the kinematic threshold the daughter pion is emitted with a very small momentum and may hence not reach the detectors. In the $\Omega^- \to \Lambda K^-$ decay, the momentum is shared more equally between the $\Lambda$ and the kaon which implies that the decay products have a larger chance to be detected also close to the production threshold. Nevertheless, the $\Omega^-$ is to date less explored than its single- and double-strange siblings. Most existing measurements of the $\Omega^-$ properties are based on small data samples of for example 4,035 events \cite{besiiiOmega, pdg}. 

The expected yields of single- and double-strange hyperons in PANDA are large, typically between 10$^4$ - 10$^6$ events per day at the initial luminosity \cite{pandahyp}. The $\Omega^-$ hyperon on the other hand has never been studied with antiproton beams before and the production cross section is unknown. The reconstruction efficiency for tracks from the full $\bar{p}p\rightarrow\bar{\Omega}^+\Omega^- \to K^+\bar{\Lambda}K^-\Lambda \to K^+\bar{p}\pi^+K^-p\pi^-$ process would provide a first estimate of the feasibility for future $\Omega^-$ hyperon studies with PANDA.

\begin{figure}[h!]
\begin{center}
\includegraphics[width=0.8\textwidth]{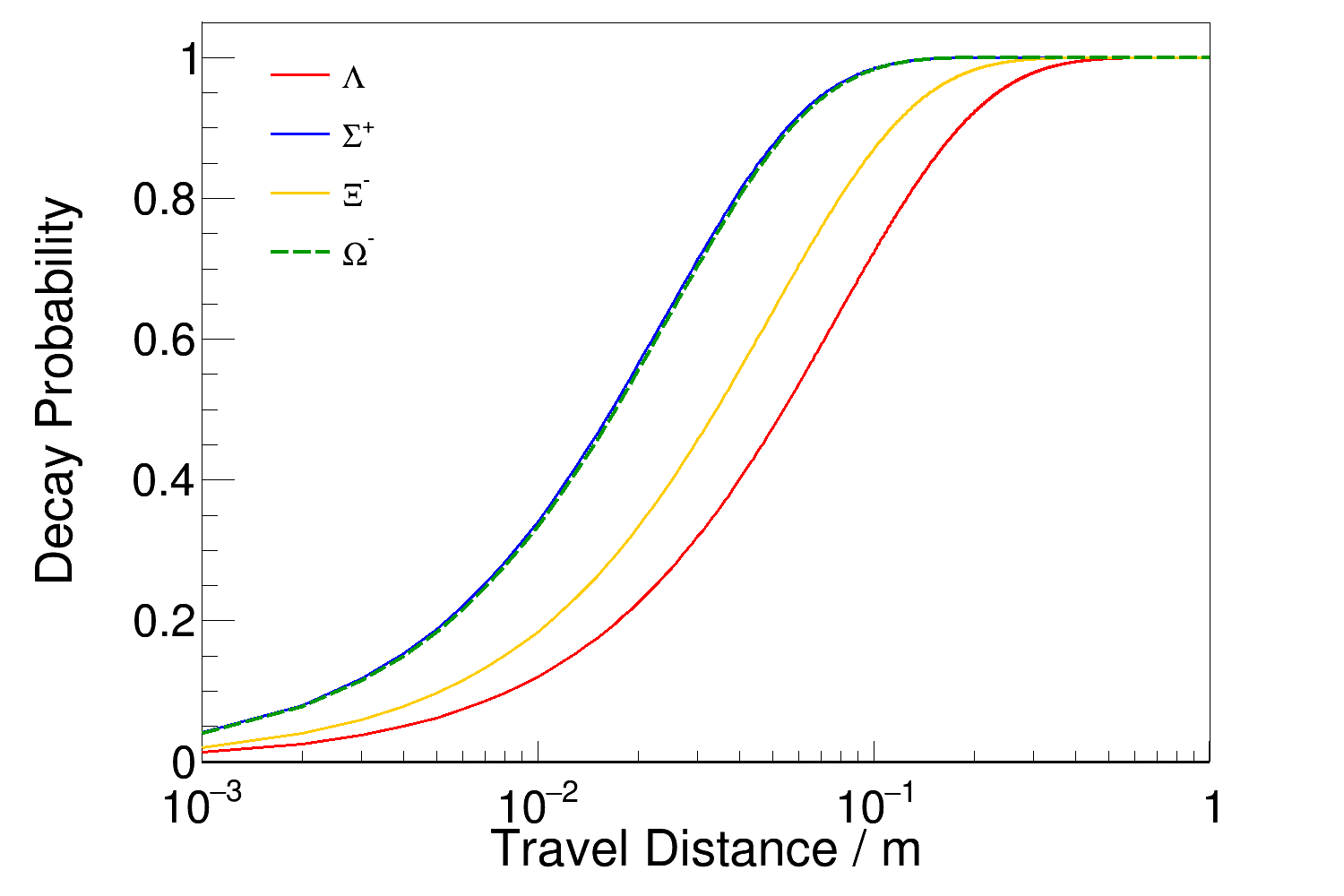}

\end{center}
\caption{The decay probability as a function of the path length for the ground-state hyperons with $c\tau$=1.}
\label{fig:hyptau}
\end{figure}

Ground-state hyperons like the $\Omega$ and its daughter, the $\Lambda$, decay weakly with a life-time of approximately 10$^{-10}$s. This means that when produced in accelerator experiments at the GeV scale, they can travel a distance of centimeters or even up to a meter before decaying. In Fig. \ref{fig:hyptau}, this is illustrated for some hyperons by the decay probability plotted against the distance traveled. The relatively long life-time gives rise to a distinct event topology with a decay vertex that is separated from the production vertex by a measurable distance. This has the advantage that it enables a very efficient suppression of background channels, but it also comes with a challenge: the major part of the standard track reconstruction algorithms on the market are optimised for particles originating at the point where the initial beam-target or beam-beam interaction takes place \cite{walterphd,PzFinder}, \textit{i.e.} the Interaction Point (IP). 

In order to maximally exploit the high interaction rates -- of up to 20 MHz -- in the PANDA experiment, an entirely software-based trigger system will be used. This will operate in real time on the data stream from the detector subsystems. Spatial information alone will not be sufficient to properly separate tracks from different events. Therefore, time information will be utilized in addition to perform \textit{4D tracking} \cite{jenny4D} where time and spatial information are combined. For tracks starting at the interaction point, tracking can be simplified enormously by constraining the IP. This reduces the amount of data and allows for fast tracking and filtering algorithms. For hyperons, this approach will however not be feasible and therefore, hyperons need customized tracking solutions.

The goal of this paper is to develop criteria for 4D tracking and triggering in the PANDA target spectrometer by:
\begin{itemize}
    \item Identifying relevant criteria describing detector signatures for specific hyperon event topologies.
    \item Evaluating these criteria for $\Omega^-$ hyperons and their decay products.
    \item Investigating which detectors yield reliable time-stamps for the reconstruction of final state particles of hyperon decays.
    \item Determining the track efficiency for different beam momenta for $\Omega^-\bar{\Omega}^+$ pairs and their decay products.
\end{itemize}

\noindent The paper is organised as follows: In Section \ref{sec:panda}, the PANDA experiment at FAIR will be presented, with emphasis on the target spectrometer and the central tracking system of PANDA. In Section \ref{sec:sim}, the simulation studies are presented. The resulting detector signatures from the $\bar{p}p \to \bar{\Omega}^+\Omega^-$ reaction are presented in Section \ref{sec:results}. In particular, we present the track reconstruction efficiency of the $\bar{p}p\rightarrow\bar{\Omega}^+\Omega^- \to K^+\bar{\Lambda}K^-\Lambda \to K^+\bar{p}\pi^+K^-p\pi^-$ reaction using the full PANDA detector as a function of the beam momentum. Finally, we discuss a possible layout of a software trigger scheme for $\bar{\Omega}^+\Omega^-$ selection in PANDA.

\section{The PANDA Experiment at FAIR}
\label{sec:panda}
\noindent The PANDA experiment, under construction at FAIR \cite{fair} in Darmstadt, Germany, will be the world's next-generation facility for studies of the strong interaction and fundamental symmetries \cite{panda}. Combining an antiproton beam in the intermediate energy range with a detector covering almost 4$\pi$ of the full solid angle provides unique conditions for hadron physics in general and hyperon physics in particular.

Antiprotons will be cooled and stored in the High Energy Storage Ring (HESR) \cite{hesr}. In the early phase of operation \cite{phaseone}, the HESR will be able to accumulate up to $10^{10}$ antiprotons in 1000~s. Stochastic cooling will result in a relative beam momentum resolution better than $\Delta p / p = 5\cdot 10^{-5}$. The antiproton beam will impinge on a hydrogen cluster jet or pellet target, yielding an average luminosity of about $10^{31}$ cm$^{-2}$s$^{-1}$~\cite{target}. This corresponds to an interaction rate of about 1 MHz.

In a later phase, the introduction of the Recuperated Experimental Storage Ring (RESR) will result in up to $10^{11}$ antiprotons to be injected and stored in the HESR. This will enable the design luminosity of about $2\cdot10^{32}$ cm$^{-2}$s$^{-1}$ to be reached, corresponding to an interaction rate of 20 MHz.

\subsection{The PANDA Detector}

\begin{figure}[h!]
\begin{center}
\includegraphics[width=0.9\textwidth]{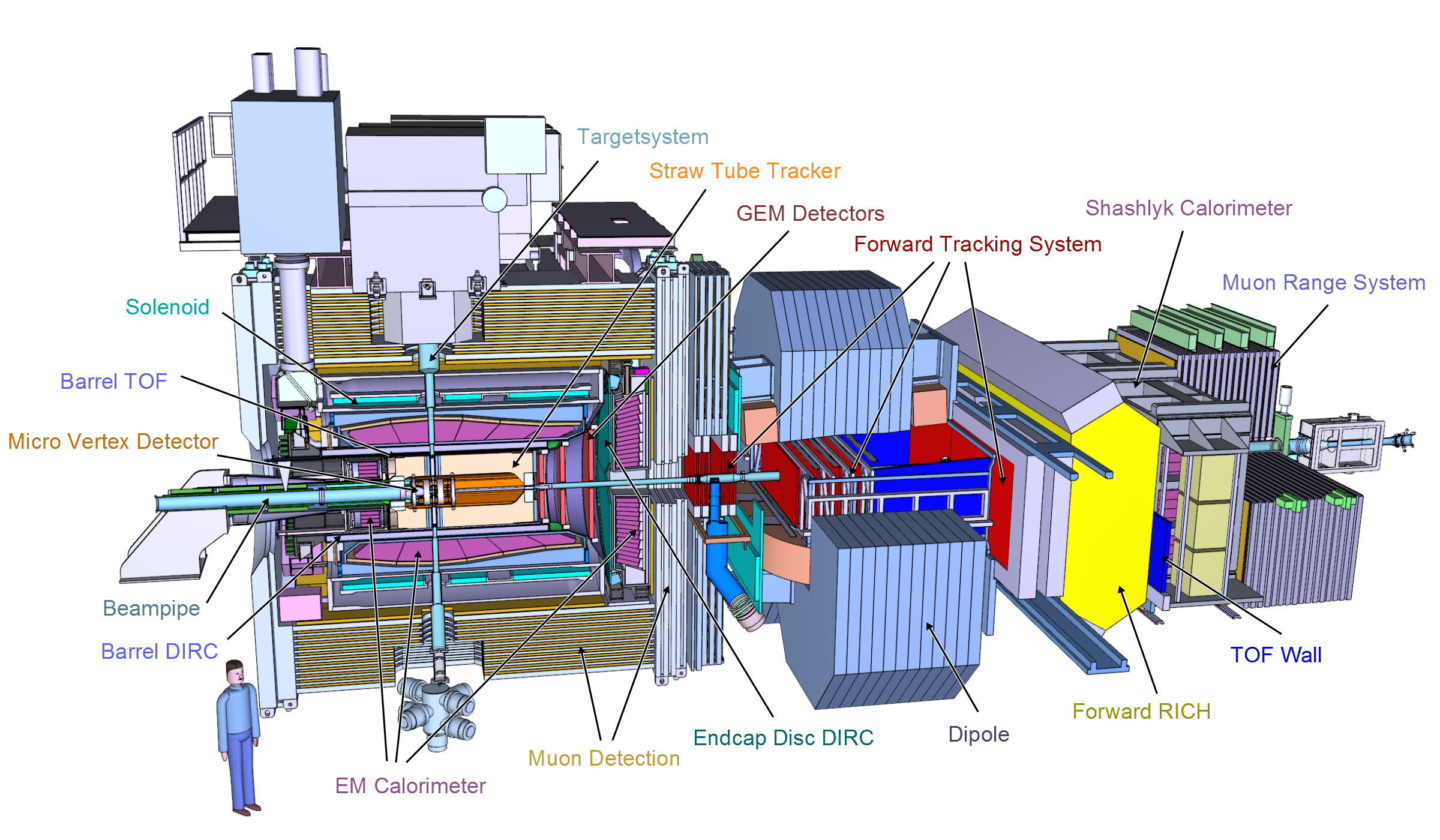}
\end{center}
\caption{The PANDA detector with the subdetectors labelled. The antiproton beam enters from the left and the cluster jets or pellet target enters from the top.}
\label{fig:panda}
\end{figure}

\noindent The PANDA detector, shown in Fig. \ref{fig:panda} and described in detail in Ref.~\cite{pandadet}, comprises two parts: the target spectrometer (TS), covering polar angles above $10^{\mathrm{o}}$ in the horizontal direction and $5^{\mathrm{o}}$ in the vertical direction, and the forward spectrometer (FS), covering angles below $10^{\mathrm{o}}$. The TS consists of the silicon Micro Vertex Detector (MVD), the gas-filled Straw Tube Tracker (STT), the gas electron multiplier detectors (GEM), the barrel Time-of-Flight detector (BarrelToF), a Detector of Internally Reflected Cherenkov light (DIRC), a lead-tungstate (PbWO$_4$) Electro Magnetic Calorimeter (EMC) and a muon detector system. A solenoid magnet, providing a field of up to 2.0 Tesla, enables measurements of charged particle momenta.

The FS will include the Forward Tracking Stations (FTS), \textit{i.e.} six straw tube stations for tracking. In addtion, it will comprise a dipole magnet, a ring imaging Cherenkov counter (RICH) detector, a forward ToF, a Shashlyk electromagnetic calorimeter and a muon range system.

The PANDA luminosity will be measured by a detector (furthest to the right in the figure)~\cite{luminosity}.

\subsubsection{The PANDA Central Tracking System}

\begin{figure}[h!]
\begin{center}
\includegraphics[width=0.7\textwidth]{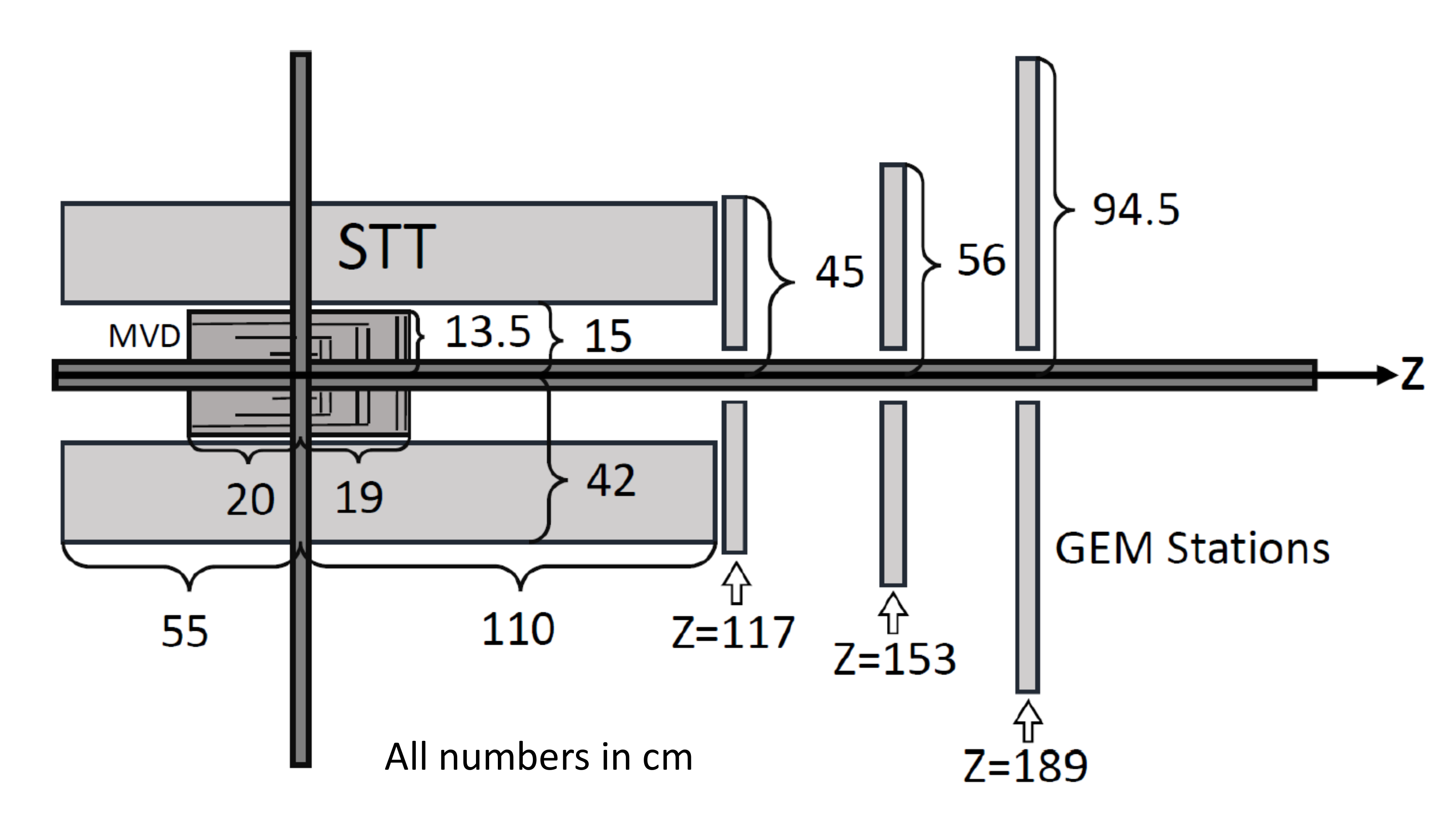}
\end{center}
\caption{Schematic layout of the MVD, STT and GEM stations. The barrel and disc parts of the MVD are indicated by black lines. All numbers represent the fiducial volume and are given in cm.}
\label{fig:SchematicMvdStt}
\end{figure}

\noindent The central tracking system, at focus in this work, consist of the MVD, STT, and GEM detectors organised according to the layout in Fig.~\ref{fig:SchematicMvdStt}. In the following, we will discuss these detectors in more detail. 

The innermost detector, surrounding the IP, is the Micro Vertex Detector (MVD)~\cite{MVD}. It consists of a barrel part and six disk layers in forward direction along the beam pipe. The purpose of the MVD is to provide hit points for all charged tracks. It will also constrain the origin of tracks from weakly decaying hadrons, \textit{e.g.} $D$ mesons and hyperons. The MVD will cover a radial region up to 13.5 cm, with the innermost and outermost barrel layers located at distances of 2.5 cm and 13.5 cm from the IP, respectively. The polar angle coverage of the MVD is from 3$^{\circ}$ up to 150$^{\circ}$. The MVD utilizes semiconductors in the form of pixels and strips. The barrel part will consist of four layers: the two innermost ones instrumented with pixels and the two outermost layers of rectangular Double Sided Silicon Strip Detectors (DSSD). The forward part comprises six disks with pixel sensors, where the two disks located most downstream along the beam pipe will be equipped with trapezoidal DSSDs. The vertex resolution of the MVD has been found to be of the order of 50 $\mu m$ in $x$-direction and $y$-direction and about 100 $\mu m$ in $z$-direction \cite{Kliemt}. Besides vertex information, the MVD will contribute to particle identification (PID) through energy loss (dE/dx) measurements and also provide time information, with a resolution of about 10 ns.

\begin{figure}[h!]
\begin{center}
\includegraphics[width=0.85\textwidth]{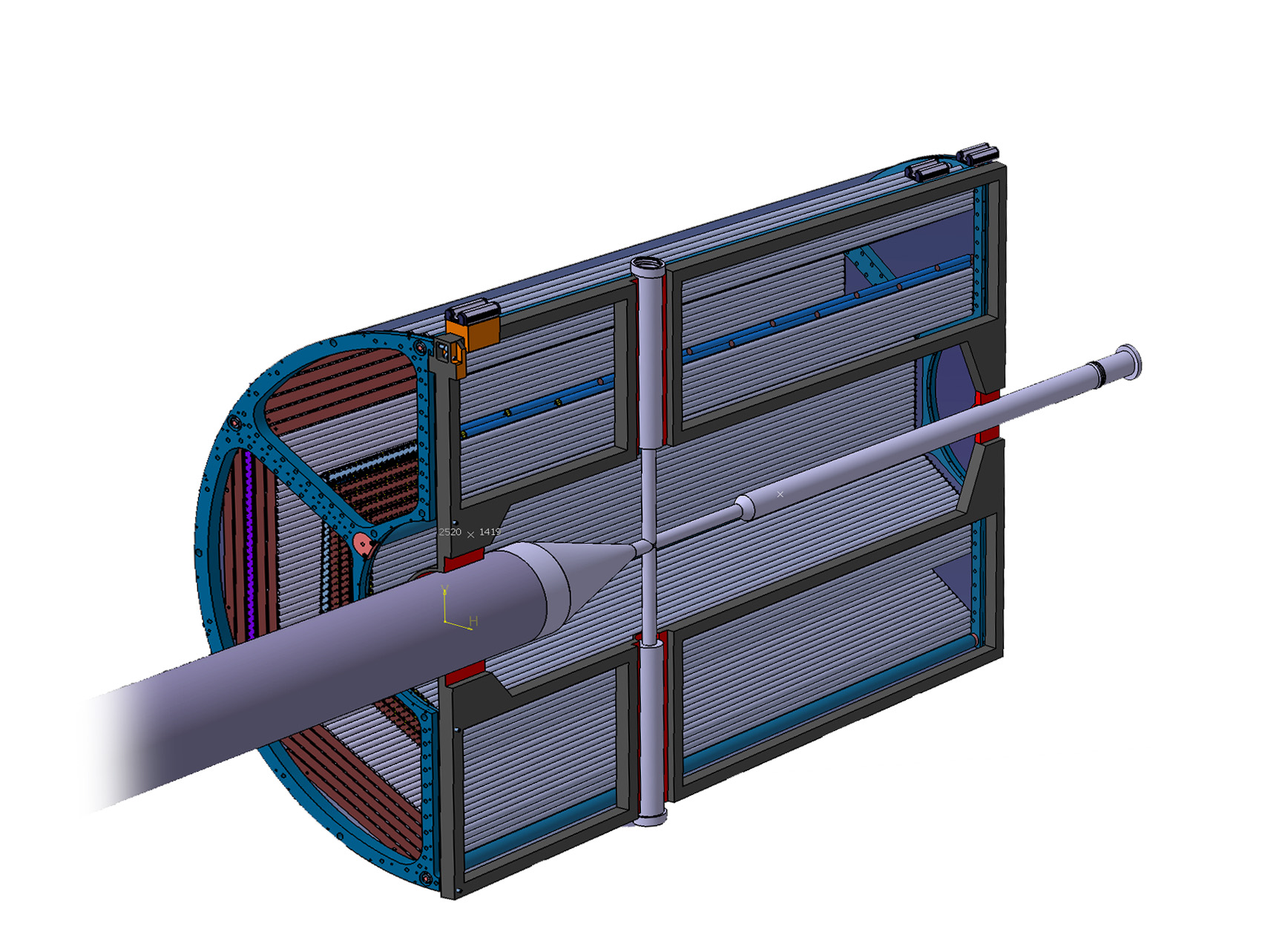}
\end{center}
\caption{Cross section of the PANDA Straw Tube Tracker (STT).}
\label{fig:STT}
\end{figure}

The Straw Tube Tracker (STT)~\cite{STT}, surrounding the MVD, will be the major component of the central tracking system. Apart from tracking of charged particles, the STT will be used for PID of charged particles with momenta below 1 GeV/\textit{c} through the energy loss (dE/dx) measurement. The STT will consist of 140 cm long straw tubes, covering a radial distance between 15.0 cm and 41.8 cm and polar angles between 10$^{\circ}$ and 140$^{\circ}$. The straws are made of aluminum-coated mylar tubes with a diameter of 10 mm with a gold-plated tungsten-rhenium anode wire in the center. They will be filled with an Argon-based gas (90$\%$) with carbon dioxide as a quench gas (10$\%$). About 4,200 drift tubes will be arranged in a closely packed hexagonal pattern as illustrated in Fig.~\ref{fig:STT}. Out of the 27 layers, 15-19 layers will be arranged parallel to the beam axis. This will provide $xy$-reconstruction in the $r\phi$-plane, where $r$ is the radial position and $\phi$ is the azimuthal angle around the beam axis. The remaining eight central layers are skewed tubes arranged with a stereo angle of $\pm3^{\circ}$ relative to the beam axis, enabling the reconstruction of the longitudinal track coordinate $z$ \cite{PzFinder}. In the $r\phi$-plane, the resolution is 150 $\mu$m. The longitudinal position resolution is 2-3 mm and the maximum drift-time in the straws is 220 ns. 

The transition region between the target and forward spectrometer will be covered by three Gas Electron Multiplier (GEM) stations. These stations, consisting of double-planes, will be located at 117 cm, 153 cm and 189 cm downstream of the IP and have a diameter of 90 cm, 112 cm and 148 cm, respectively. The GEM will cover polar angles between 3$^{\circ}$ and 20$^{\circ}$. Each station will have double-sided readout, meaning that a particle traversing both the front and back of a plane will give rise to two signals.

The tracking of charged particles can be further improved by a precise reference time, $t_0$, from a fast detector. For this purpose, and for providing PID information, the BarrelToF detector~\cite{BTOF} has been designed. It will surround the STT at a radial distance of roughly 50 cm from the beam pipe and cover polar angles between 22$^{\circ}$ and 140$^{\circ}$. The BarrelToF will consist of 1,920 scintillator tiles of size 87.0$\times$29.4$\times$5.0 mm$^3$. For particles hitting the centre of a tile, a time resolution of 55.5 ps has been achieved~\cite{Zimmermann}. 

\subsection{Data Readout and Triggering}

\noindent The PANDA data will be generated and read-out continuously, or \textit{free-streaming}, and processed to form tracks and clusters which in turn are combined into events. The software trigger then tests whether the events fulfill the requirements of the trigger lines and if this is the case, the event is stored for future analysis. The event building and trigger decisions will be carried out in real time during the operation of the experiment. The required reduction factor of the full trigger stream must be around 1,000, resulting in a maximum event rate of 20 kHz being passed to the permanent storage.

At large beam intensities, subdetector signals from different events will overlap in time. Assuming that the time intervals between interactions are Poisson distributed, the mean time between events at 2 MHz is 500 ns while 20 MHz at 50 ns. Given a maximum STT drift time of 220 ns, the 2 MHz does not lead to much overlap, while 20 MHz does. In the latter case, spatial information alone is not sufficient to reconstruct tracks and events. However, by including time information, signals can be correctly combined using tracking algorithms that are suitable for free-streaming data.

The MVD, GEM and BarrelToF detectors all operate on shorter time scales (between 10 ns and 55 ps) and provide further constraints for track and event building when combining the information with those from the STT.

\subsection{The PANDA Software}

\noindent The simulation studies presented in this work are performed within the framework PandaROOT \cite{PANDAROOT}. The full simulation chain includes Monte Carlo event generation, particle propagation and simulation of the particle's response, hardware digitization. As further steps, reconstruction and data analysis is performed. The last three steps are common for experimental and simulated data. PandaROOT is derived from the FairROOT framework \cite{FAIRROOT} which in turn is based on ROOT \cite{ROOT}.

\section{Hyperon Detector Signatures}
\label{sec:sim}
\begin{figure}[h!]
\begin{center}
\includegraphics[width=0.6\textwidth]{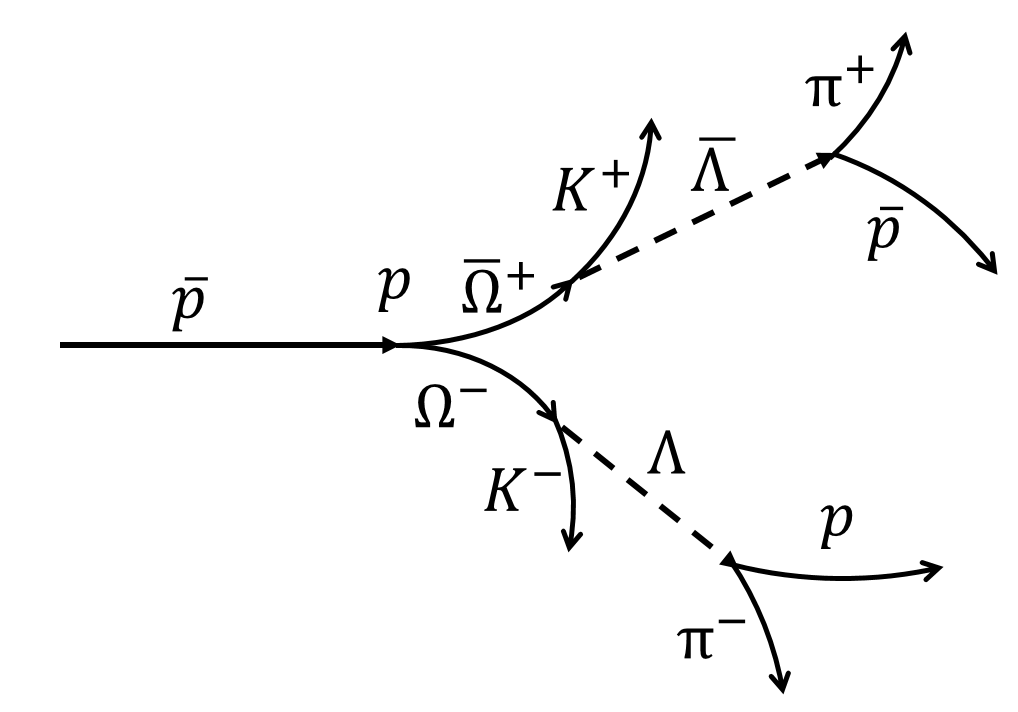}
\end{center}
\caption{Illustration of the $\bar{p}p\rightarrow\bar{\Omega}^+\Omega^-\rightarrow K^+\bar{\Lambda}K^-\Lambda\rightarrow K^+\bar{p}\pi^+K^-p\pi^-$ reaction. Solid lines correspond to charged particles and dashed lines correspond to neutral particles.}
\label{fig:OmegaDecay}
\end{figure}

\noindent The reaction studied in this work is $\bar{p}p\rightarrow\bar{\Omega}^+\Omega^- \to K^+\bar{\Lambda}K^-\Lambda \to K^+\bar{p}\pi^+K^-p\pi^-$ at beam momenta of $4.96$ GeV/$c$, $5.0$ GeV/$c$, $7.0$ GeV/$c$ and $15.0$ GeV/$c$. The $4.96$ GeV/$c$ and $5.0$ GeV/$c$ points are close to the kinematic threshold -- the excess energies are 1.2 MeV and 17.7 MeV, respectively -- and therefore represent the low energy extreme. The 5.0 GeV/\textit{c} point has been studied in some detail while at 4.96 GeV/$c$, only the efficiency has been evaluated. These two points cover the edges of an energy region where interaction studies are possible. The $7.0$ GeV/$c$ point coincides with a planned campaign for measuring the line-shape of $X$(3872) \cite{xscan}, while the $15.0$ GeV/$c$ point is chosen because it is the largest momentum that the HESR will offer. In this sense, it constitutes the high energy extreme of the experiment.

The event topology of this reaction is shown in Fig.~\ref{fig:OmegaDecay}. The $\Lambda$ and $\Omega^-$ hyperons decay weakly and hence at a measurable distance from the point of production. Therefore, the full decay topology includes four displaced vertices. The neutral hyperons, represented by dashed lines in the figure, do not interact with the detector material through ionization and can therefore not be tracked. Furthermore, their trajectories do not bend in the magnetic field. The $\Omega^-$ and $\bar{\Omega}^+$ are charged, their trajectories bend in the magnetic field and they decay sequentially, \textit{i.e.} to intermediate $\Lambda$ and $\bar{\Lambda}$, which in turn decay into $p\pi^-$ and $\bar{p}\pi^+$, respectively. An isotropic distribution is used for all decays because all valence quarks are annihilated and with increasing number of annihilated \textit{s} quarks the angular distribution gets more isostropic.

\subsection{Simulation Conditions}
\label{sec:simcond}

\noindent The following simplifications have been made for the simulations and reconstruction:

\noindent \textbf{100$\%$ efficient tracking:} An algorithm is used, where detector hits corresponding to those from an original \textit{true} Monte Carlo track, are grouped together into a reconstructed track.  

\noindent \textbf{100$\%$ efficient PID:} For the purpose of this study, other effects than those of detector hits should be minimized. Ideal PID, where each track is matched with the true Monte Carlo track, is therefore used. 

\noindent \textbf{Point-like target:} In the real experiment, both the cluster jet target and the pellet target will give rise to residual gas in the beam pipe. This will lead to gas particles outside of the interaction point acting as an effective target in the beam pipe close to the target region. In the simulations presented here, this effect has not been taken into account but instead, a point-like target is simulated. A study of the effect of a more realistic target has been carried out \cite{adeel}. The preliminary results reveal no significant difference in the reconstruction efficiencies between the ideal and realistic gas in the target for hyperon events originating from beam-target interactions.

\subsection{Quantities of Interest}
\label{sec:qoi}

\noindent In order to formulate criteria for software trigger lines, data reconstruction algorithms or analysis methods, we have identified quantities on the basis of the detector signatures of a given reaction. These quantities can also give an in-depth understanding of the feasibility of studying a given reaction at a given experiment. The quantities of interest are:

\begin{enumerate}
    \item Kinematic distributions
    \item Decay vertex position
    \item Tracking detector hits
    \item Detector hits for time information
    \item Track efficiency
\end{enumerate}

\noindent In the following, we will discuss these quantities in some detail.

\subsubsection{Kinematic distributions} 
\noindent The momentum of each particle in the laboratory system \textit{versus} the laboratory polar angle $\theta$, defined with respect to the beam axis, provides a first estimate the acceptance of different subdetectors. It also reveals how well the final state particles can be separated kinematically. In addition, the transverse momentum $p_T$ [GeV/\textit{c}] is of relevance because in the TS solenoid field $B$ [T], charged particles will follow helix trajectories with radius 
\begin{equation}
r = \frac{p_{T}}{|q|B}.
\label{radius}
\end{equation}
Here, the charge, $q$ [C], determines the rotational direction of the trajectory. This means that particles produced at the IP with a transverse momentum smaller than 270 MeV/c will be trapped in a spiral trajectory inside the STT. These particles will not give rise to signals in detectors outside the STT such as the BarrelToF. Particles from displaced decay vertices can however reach outside of the STT even at smaller transverse momenta.

\subsubsection{Decay vertex position} 
\noindent The decay vertex position longitudinal ($z$) and transversal ($t$) to the beam indicate in which detectors we expect signals from the decay products.

\subsubsection{Tracking detector hits} 
\noindent Whether a particle will give rise to signals from which a complete track can be reconstructed, depends on the number of hits in the main tracking detectors. In the TS, the STT and the GEM are the relevant systems and in the FS it is the FTS. We define three main categories:
\begin{itemize}
    \item A particle gives rise to sufficiently many hits to form a reconstructable track ($\geq$ 4 STT hits)
    \item A particle gives rise to too few hits to be reconstructable (0-3 STT hits)
    \item A particle gives rise to a very large number of hits ($\geq$ 50 STT hits)
\end{itemize}

\subsubsection{Detector hits for time information} 
\noindent Straw tube trackers such as the STT and the FTS rely on a precise reference time, $t_0$, to extract the drift time and thereby precisely determine the particle trajectory. The barrel and forward ToF are sufficiently fast to provide such time-stamps. Events without timestamps from these, can instead use information from the MVD or GEM for this purpose. At low luminosities, where the signals from different events will be fairly well separated in time, it will be sufficient to use one particle per event for a reasonable estimate of $t_0$. This is referred to as \textit{events with hits}.

However, at higher luminosities, signals from different events will overlap in time. Therefore, a hit in a fast detector needs to be combined with hits from the same particle in the tracking detectors. That requires that each particle has a precise time-stamp, \textit{i.e.} that each track has a corresponding hit in the MVD, GEM or Barrel ToF. This we refer to as \textit{tracks with hits}. 

\subsubsection{Track efficiency} 
\noindent A \textit{reconstructable} track has a reasonable chance to be correctly reconstructed by the pattern recognition algorithm in the TS if it contains at least four MVD hits or at least six hits in the MVD+STT+GEM and at least four STT hits. 

In order to reconstruct the $\bar{p}p \to \bar{\Omega}^+\Omega^-$ reaction exclusively, all final state tracks need to be reconstructed. To determine the efficiency, we take the full PANDA detector into account, \textit{i.e.} the TS and the FS. The efficiency is defined as the fraction of events where all final state particles give rise to reconstructable tracks with at least four MVD or six MVD+STT+GEM hits or six hits in forward tracker planes.

\section{Results}
\label{sec:results}

\subsection{Kinematic distributions}
\label{sec:kin}

\begin{figure}[h!]
\includegraphics[width=.32\textwidth]{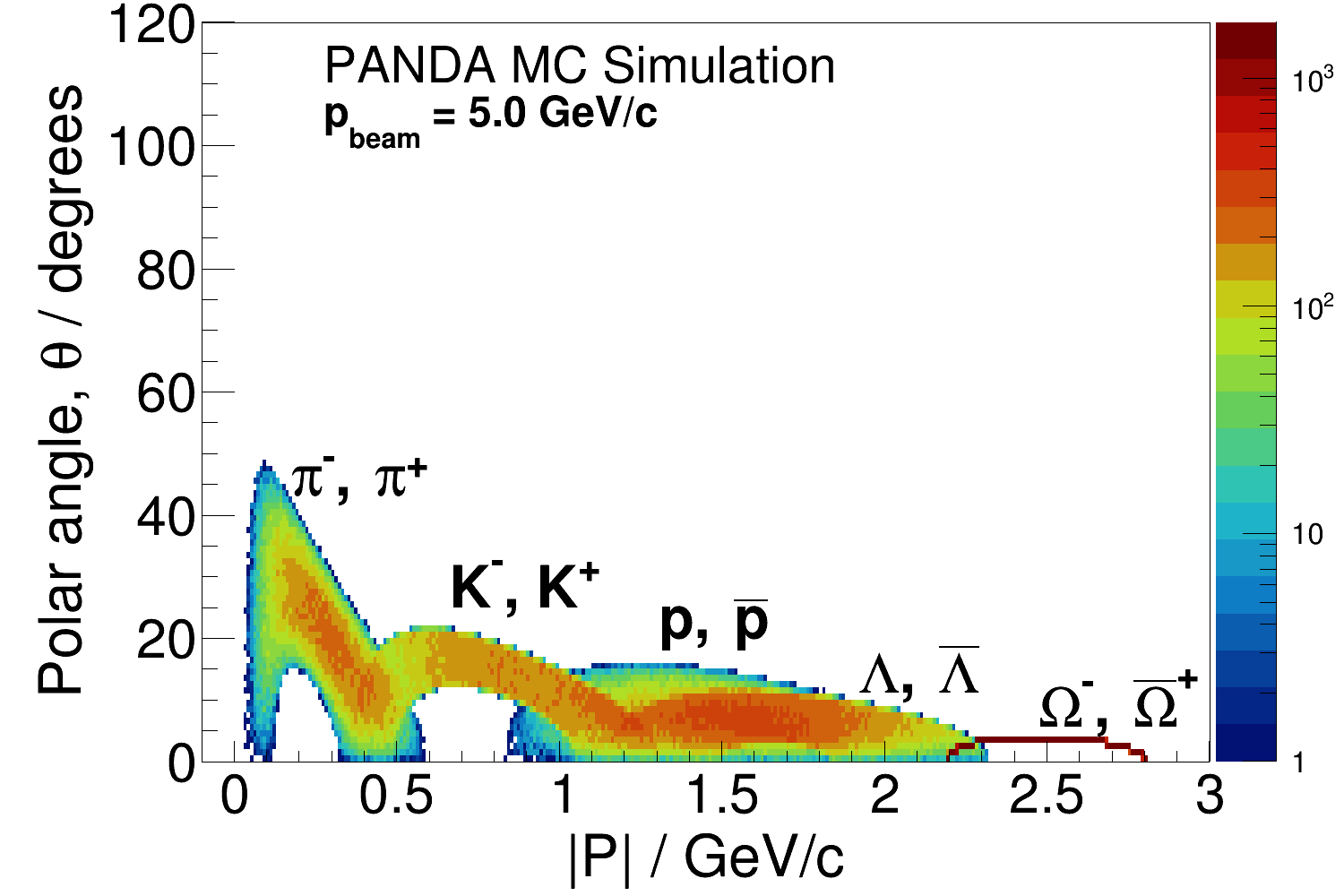}
\includegraphics[width=.32\textwidth]{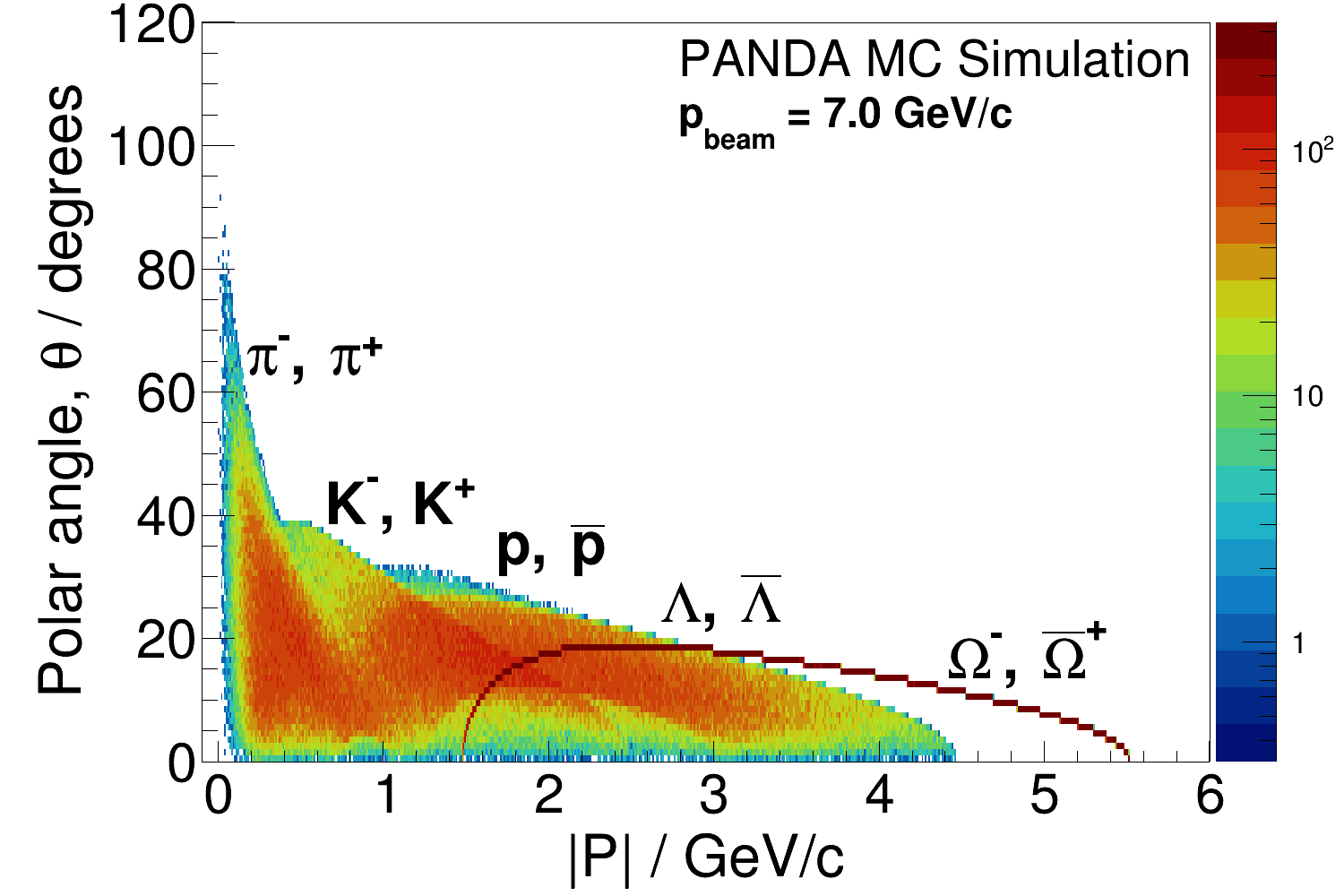}
\includegraphics[width=.32\textwidth]{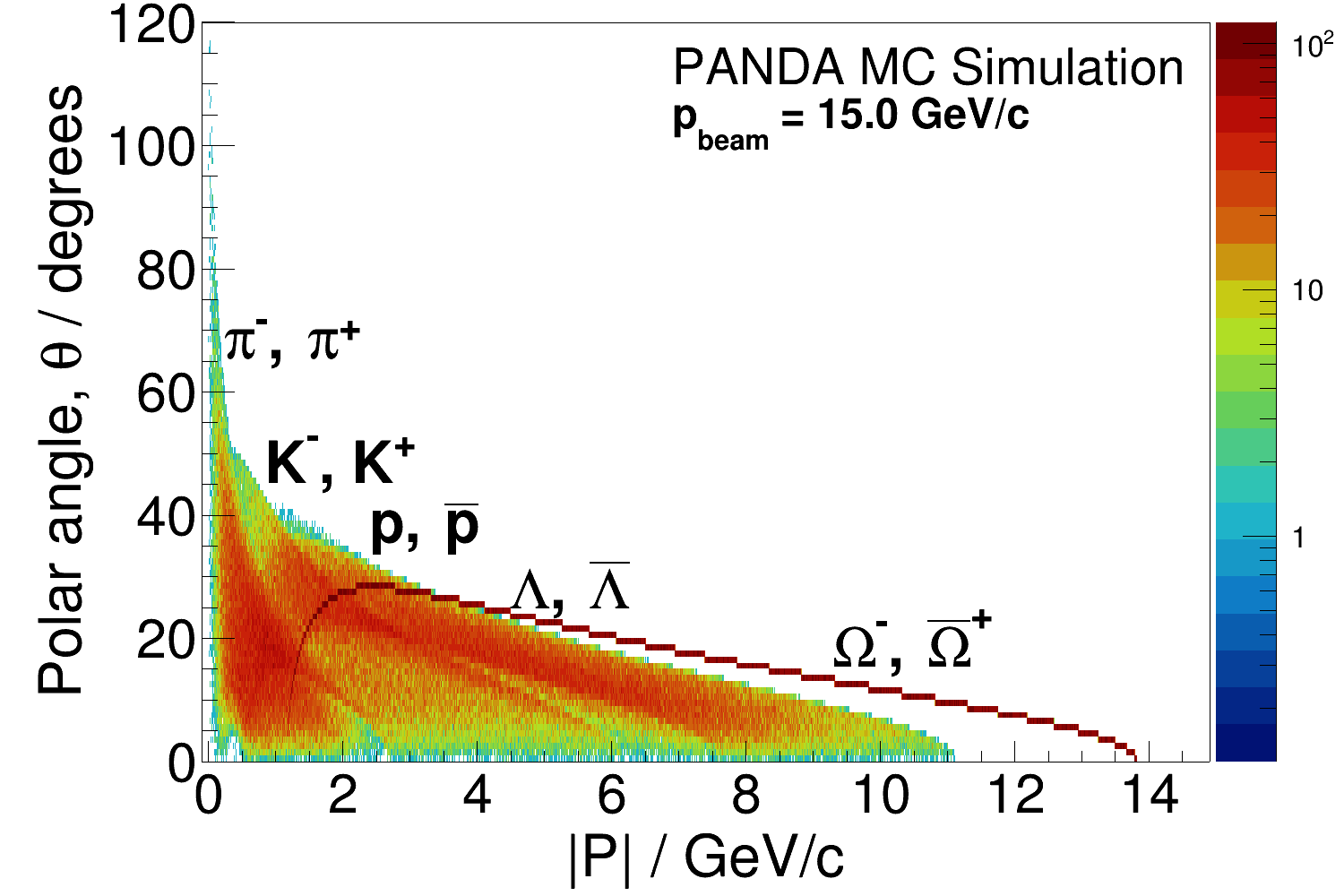}
\caption{Absolute momentum \textit{versus} the lab polar angle distributions for particles from the $\bar{p}p\rightarrow\bar{\Omega}^+\Omega^-$ reaction and its subsequent decays at beam momenta of 5.0, 7.0 and 15 GeV/\textit{c}.}
\label{fig:momentumAngle}
\end{figure}

\noindent The lab momentum \textit{versus} the lab polar angle distribution is shown for all involved particles in Fig. \ref{fig:momentumAngle}. As expected, light particles are generally emitted at larger angles and smaller momenta than heavier particles. At a beam momentum of 5.0 GeV/\textit{c}, the distributions from pions, kaons and protons/antiprotons are fairly well separated in contrast to larger beam momenta.

\begin{figure}[h!]
\includegraphics[width=.32\textwidth]{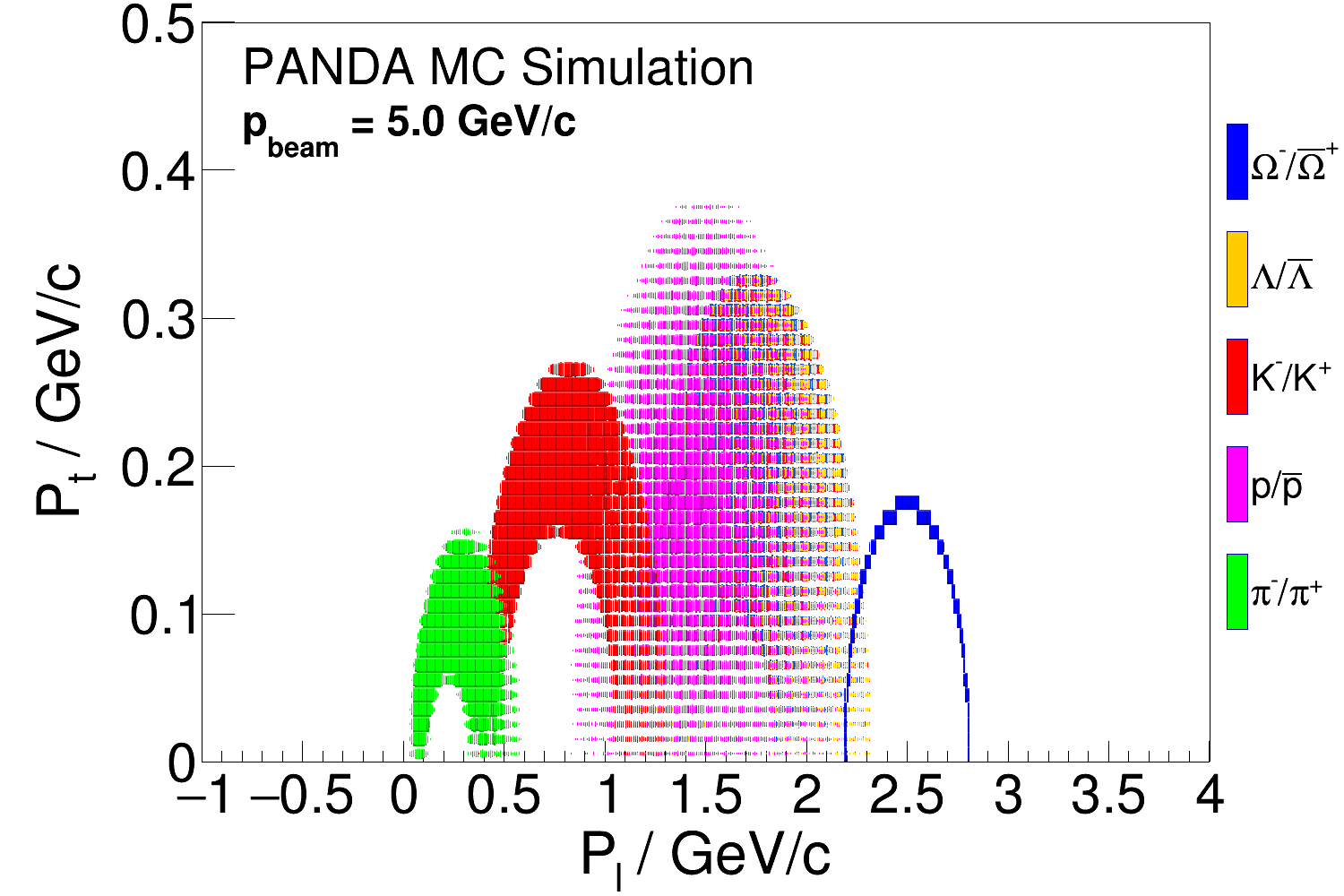}
\includegraphics[width=.32\textwidth]{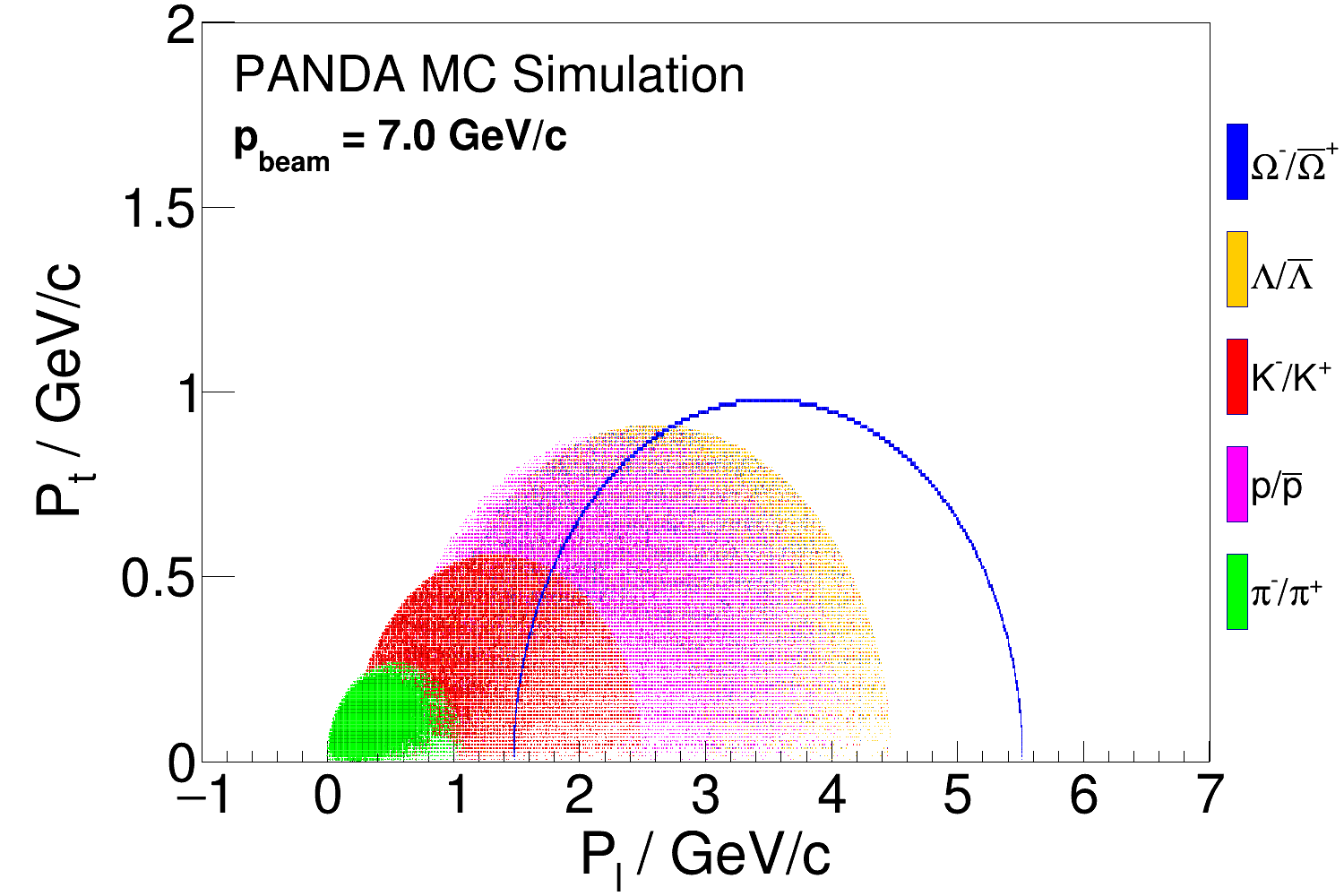}
\includegraphics[width=.32\textwidth]{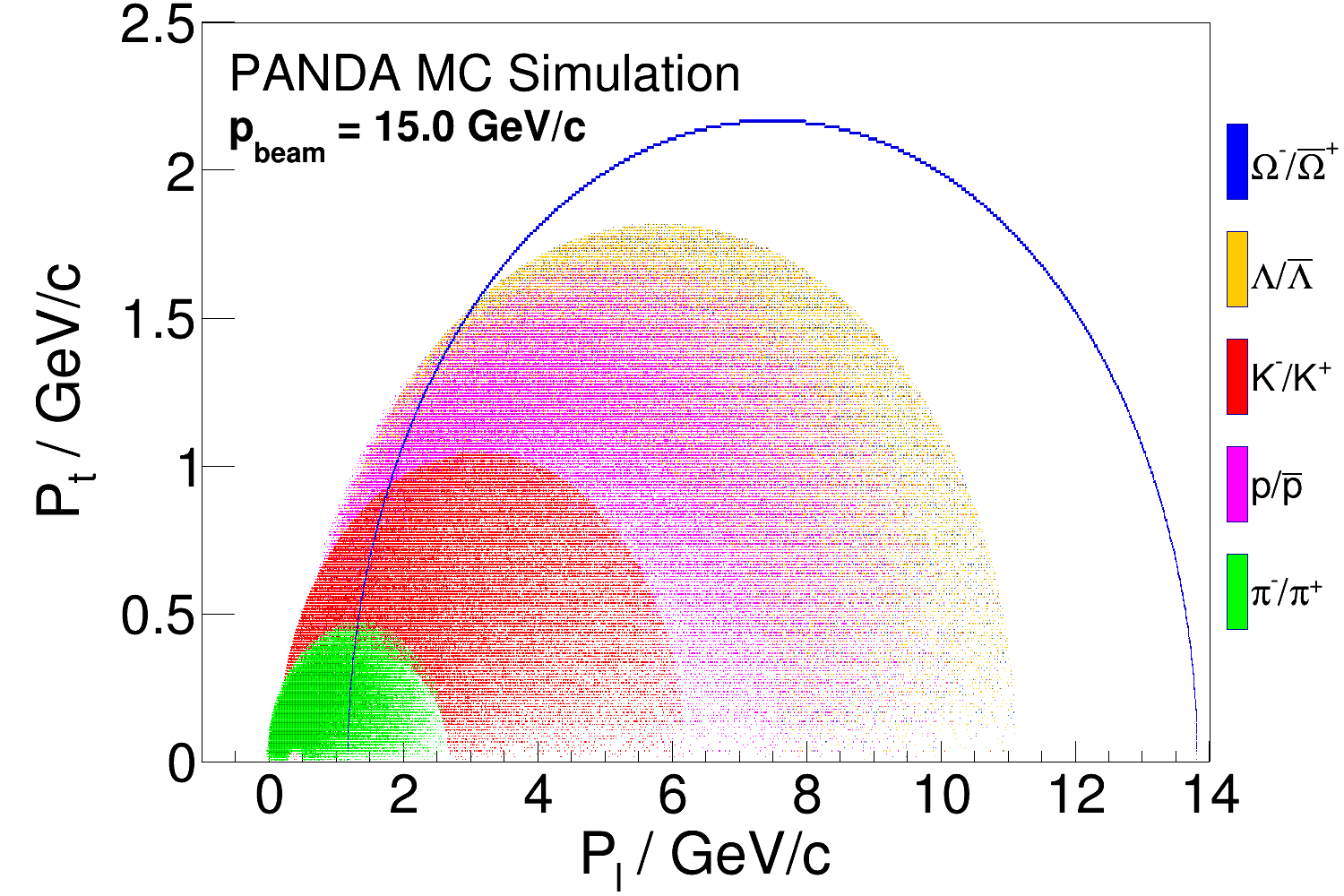}
\caption{Transverse \textit{versus} longitudinal momentum distributions for particles from the $\bar{p}p\rightarrow\bar{\Omega}^+\Omega^-$ reaction and its subsequent decays.}
\label{fig:2DMomentum}
\end{figure}

Fig. \ref{fig:2DMomentum} displays the transverse \textit{versus} the longitudinal momentum distribution of the final state and intermediate particles. At a beam momentum of 5.0 and 7.0 GeV/\textit{c} most pions have small transverse momenta. However, since they come from a displaced hyperon decay vertex, there is a possibility that they leave the STT.

\subsection{Decay Vertex Distribution}
\label{sec:vertex}

\begin{figure}[h!]
\includegraphics[width=.49\textwidth]{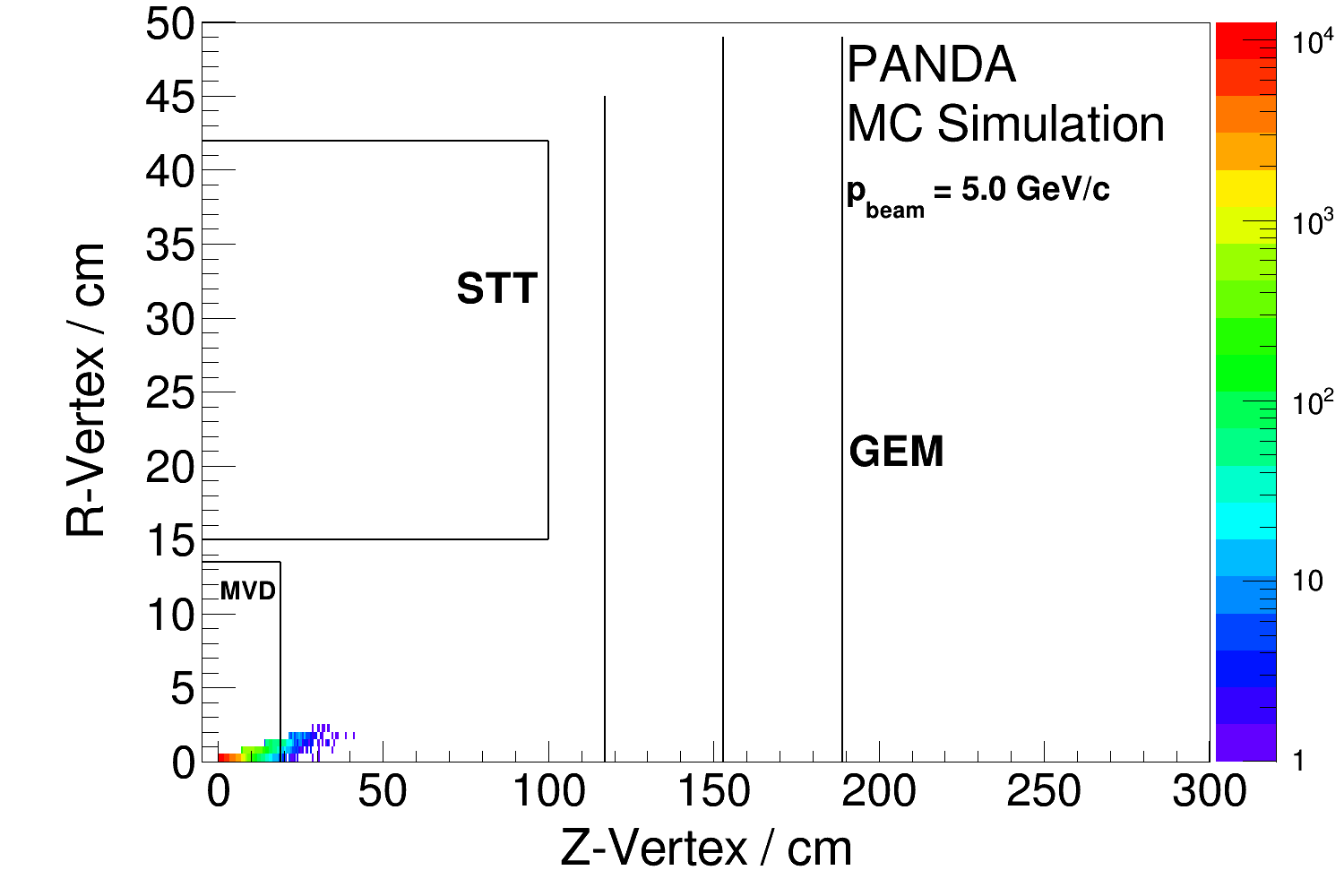}
\includegraphics[width=.49\textwidth]{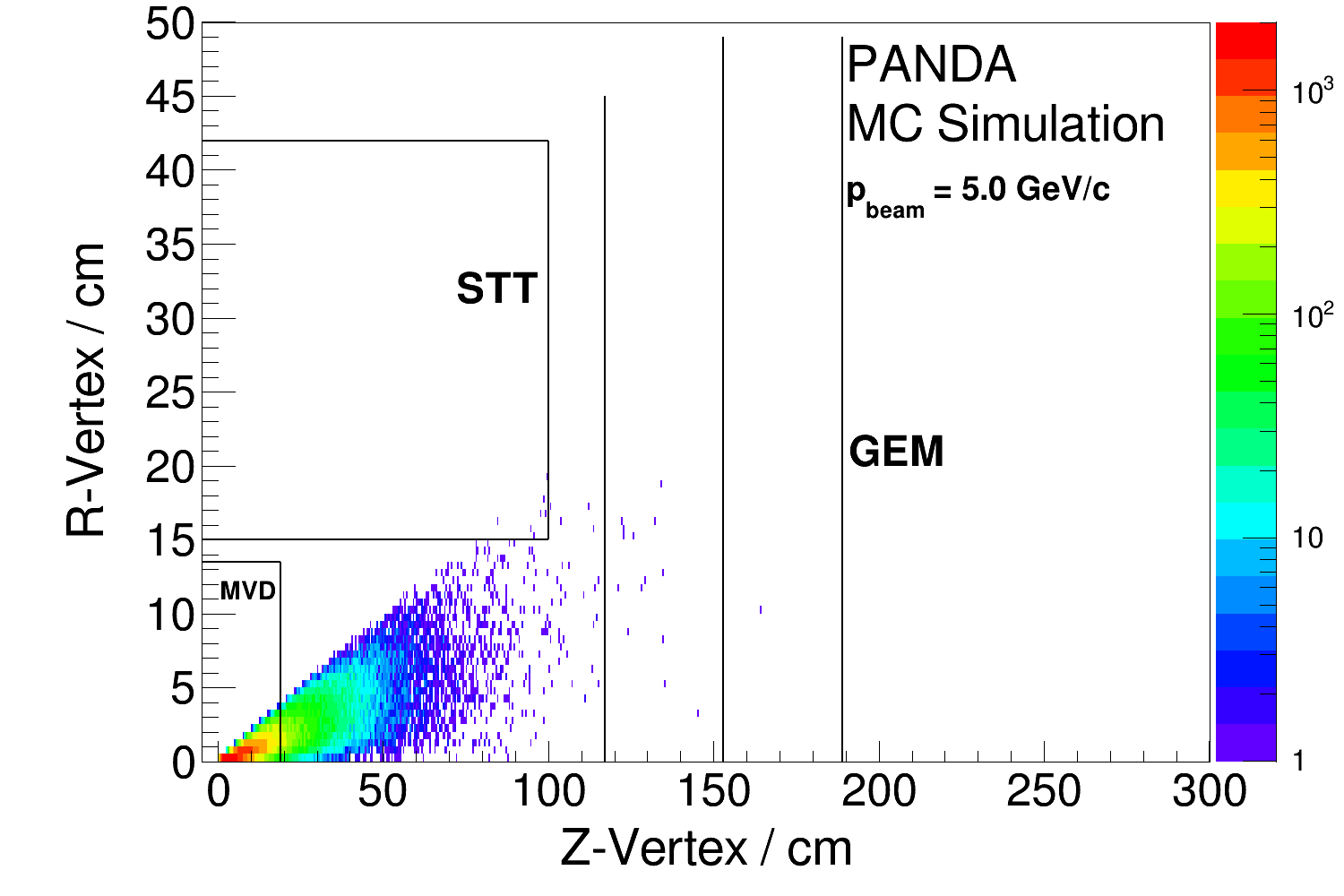}\\
\includegraphics[width=.49\textwidth]{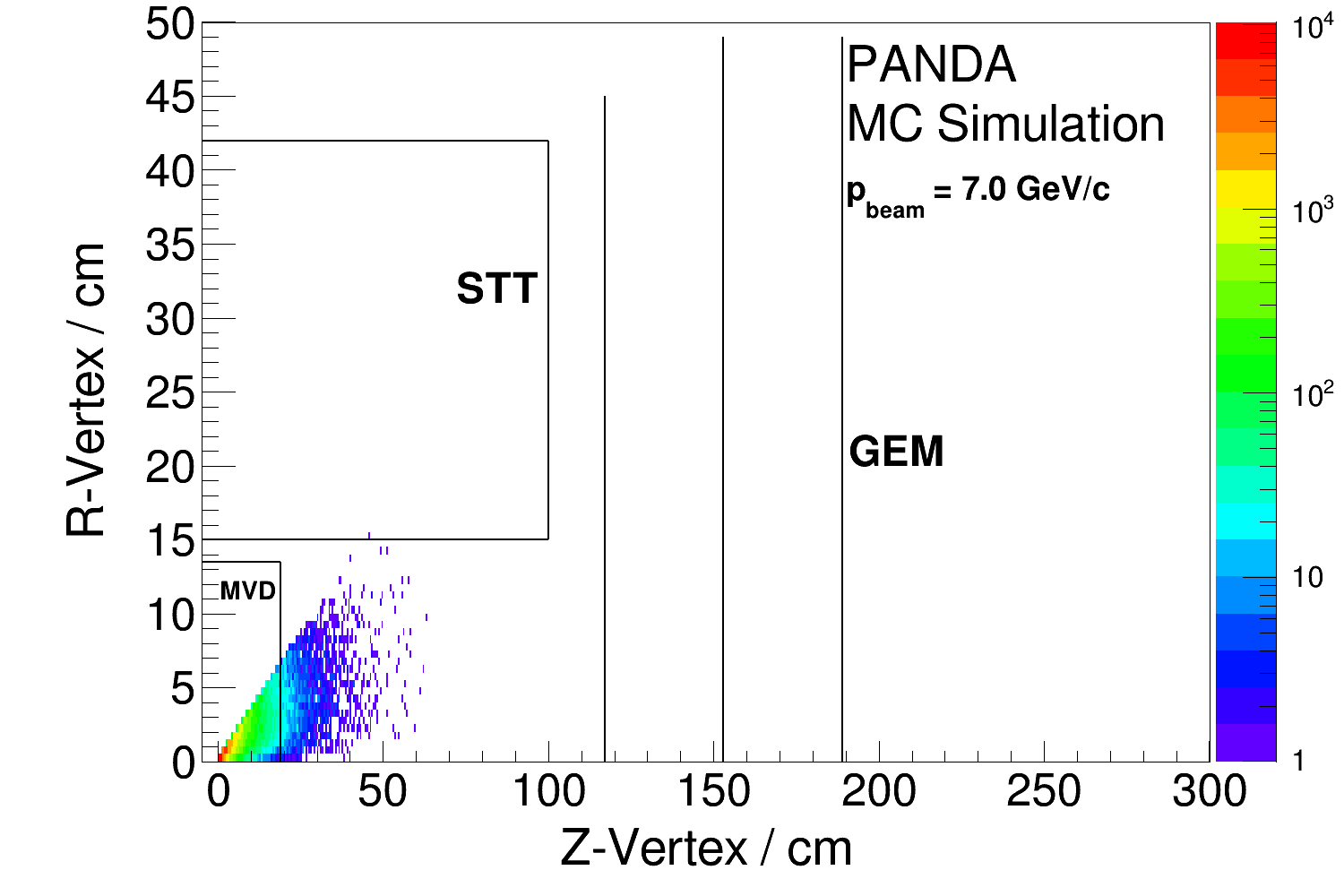}
\includegraphics[width=.49\textwidth]{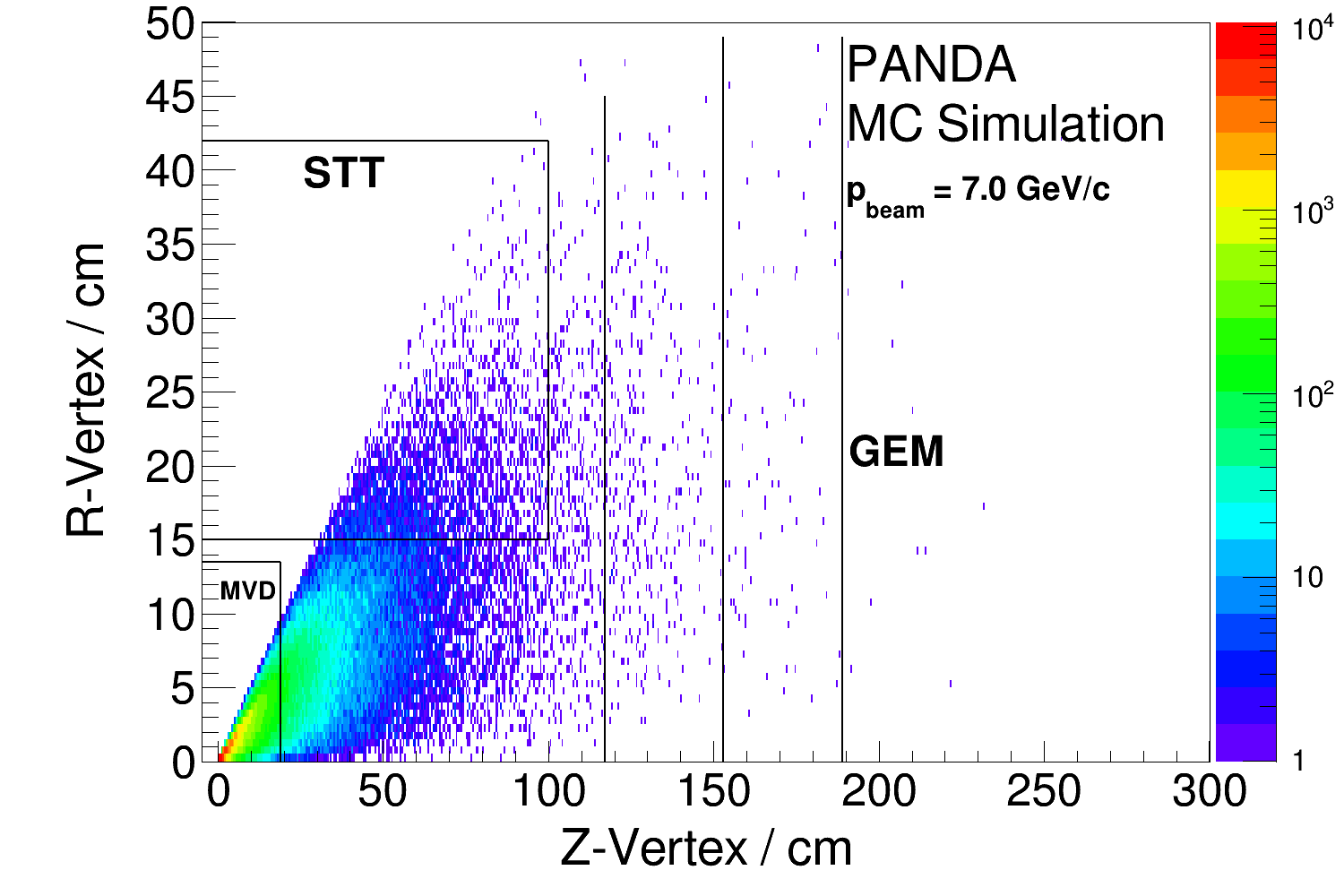}\\
\includegraphics[width=.49\textwidth]{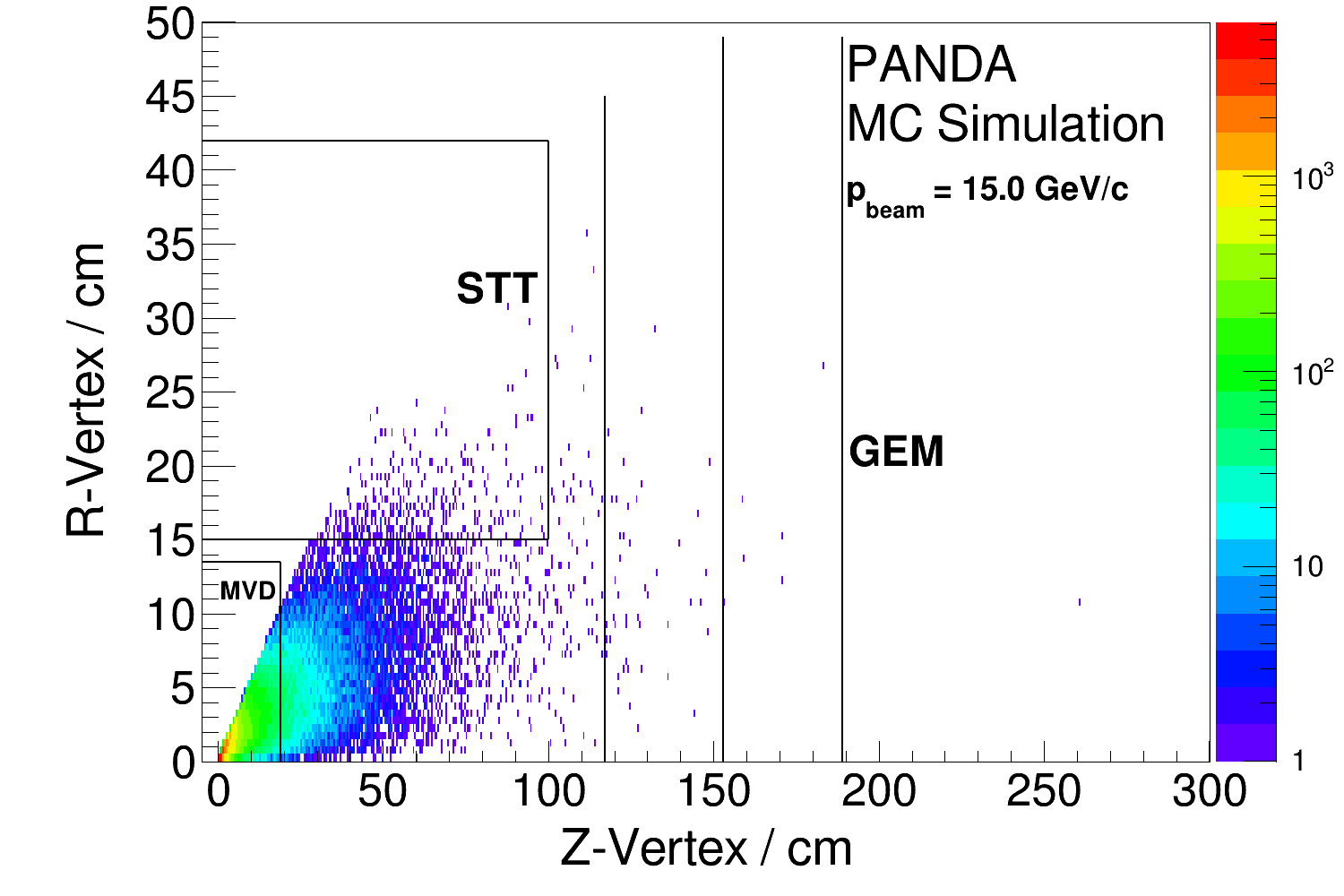}
\includegraphics[width=.49\textwidth]{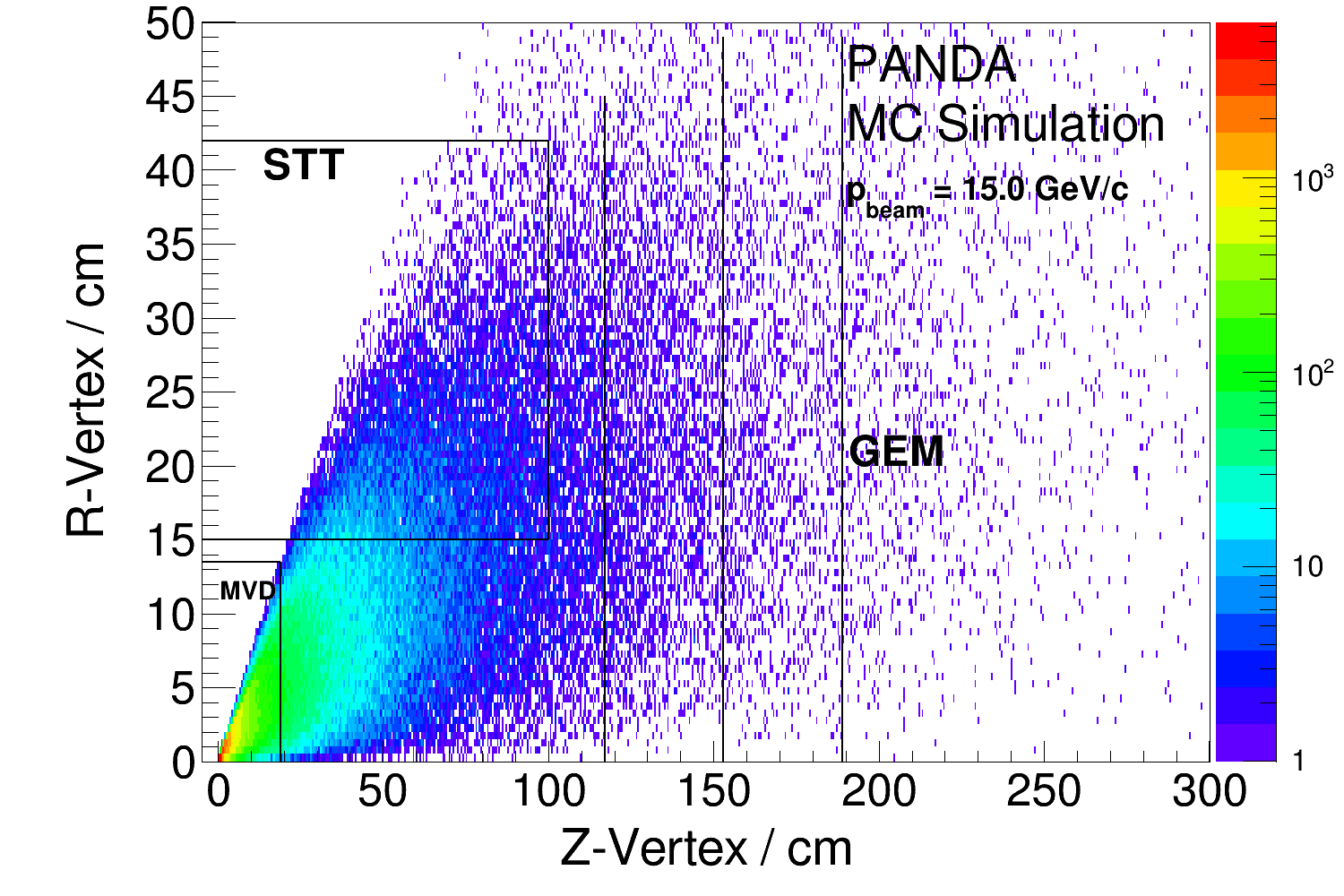}
\caption{Decay vertex positions of $\Omega^-$ and $\bar{\Omega}^+$ (left) and the $\Lambda$ and $\bar{\Lambda}$ (right). The top panels show the distributions at a beam momentum of 5.0 GeV/\textit{c}, middle panels at 7.0 GeV/\textit{c} and the bottom panel at 15.0 GeV/\textit{c}. The inner rectangle mark the boundaries of the MVD and the outer rectangle mark the boundaries of the STT. The GEM planes are marked with vertical lines.}
\label{fig:DecayVertex}
\end{figure}

\noindent The decay vertex distributions of the $\Omega$ and $\Lambda$ hyperons are shown in Fig. \ref{fig:DecayVertex} and also summarized in Table \ref{tab:decayVertexOmega}. We conclude that most $\Omega^- \to \Lambda K^-$ decays occur inside the MVD at all beam momenta, which means that the kaons give rise to signals in the MVD. Also the daughter $\Lambda$ decay to a large extent within the MVD. However, at larger beam momenta, the probability for the decay to occur between the MVD and the GEM planes increases. The decay products are then expected to leave hits in the STT, GEM or the FS. 

\begin{table}[h!]
\caption{The fraction of decay vertices within the STT and the MVD for $\Omega^-$, $\bar{\Omega}^+$, $\Lambda$ and $\bar{\Lambda}$ from the $\bar{p}p \to \bar{\Omega}^+\Omega^-$ reaction. A particle is in the TS acceptance if it has given rise to at least four MVD or at least six MVD+STT+GEM hits. Particles that do not decay in the MVD or STT volume decays outside the active detector material. The target gap is included in the acceptance.}
\begin{center}
\begin{tabular}{ | l | c | c | c |}
\hline
$\Omega^-$ and $\bar{\Omega}^+$ & 5.0 GeV/\textit{c} & 7.0 GeV/\textit{c} & 15 GeV/\textit{c}\\
\hline
MVD & 99.3\% & 96.7\% & 82.2\% \\
STT & 0\% & 0.001\% & 0.37\%\\
TS acceptance & 100\% & 100\% & 99.9\% \\
\hline
$\Lambda$ and $\bar{\Lambda}$ & & & \\
\hline
MVD & 70.9\% & 57.4\% & 32.3\% \\
STT & 0.009\% & 3.41\% & 15.9\% \\
TS acceptance & 99.97\% & 99.3\% & 90.5\% \\
\hline
\end{tabular}
\end{center}
\label{tab:decayVertexOmega}
\end{table}

\subsection{Reconstructable tracks}
\label{sec:recotracks}

\noindent The fraction of events where a given final state particle gives rise to a reconstructable track is shown in Table \ref{tab:ResultsTracksTS}. At all beam momenta, most final state particles are reconstructable in the TS. Many of the remaining particles may be reconstructable in the FS, as we will come back to in section \ref{sec:acc}. 

\begin{figure}[h!]
\begin{center}
\includegraphics[width=0.32\textwidth]{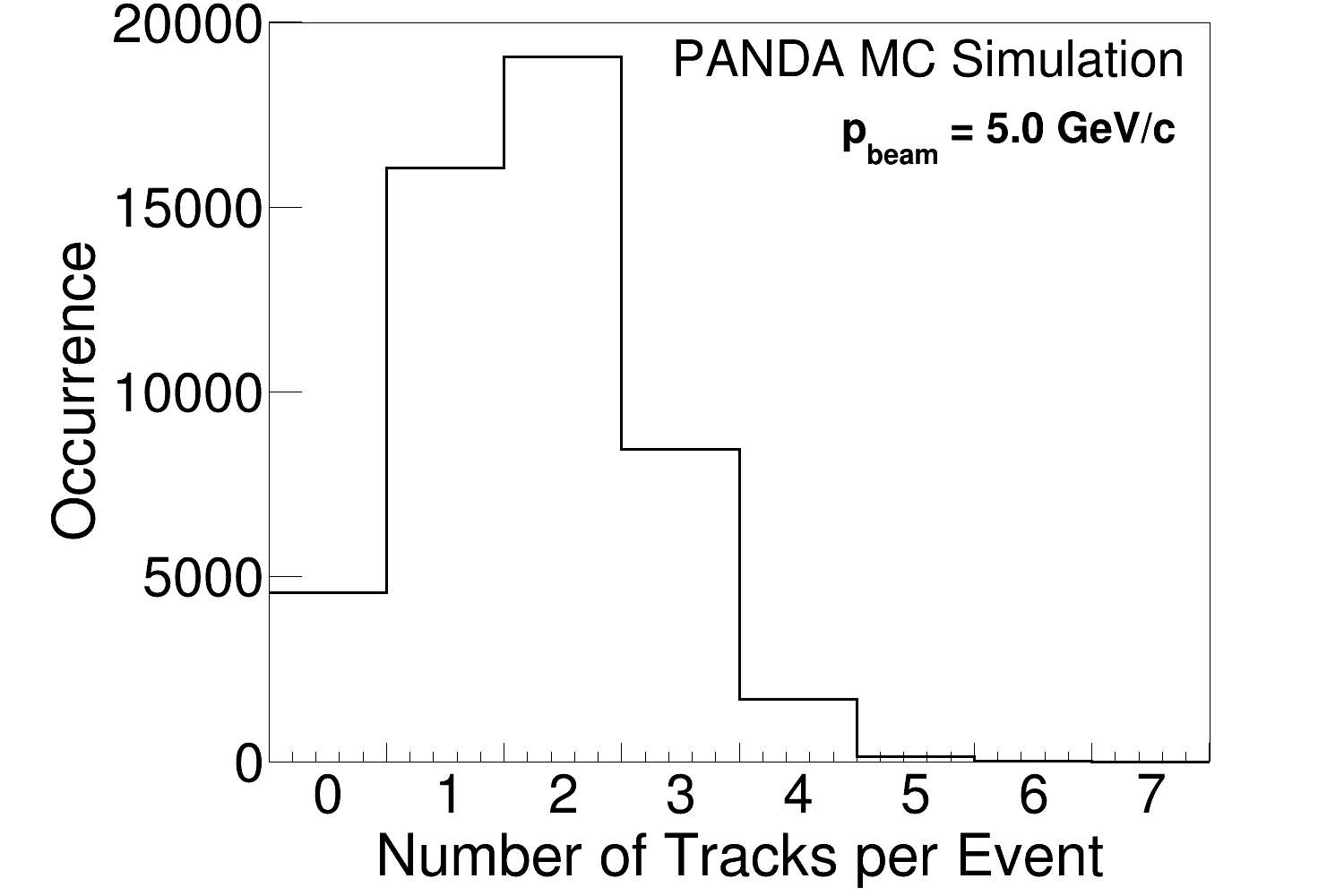}
\includegraphics[width=0.32\textwidth]{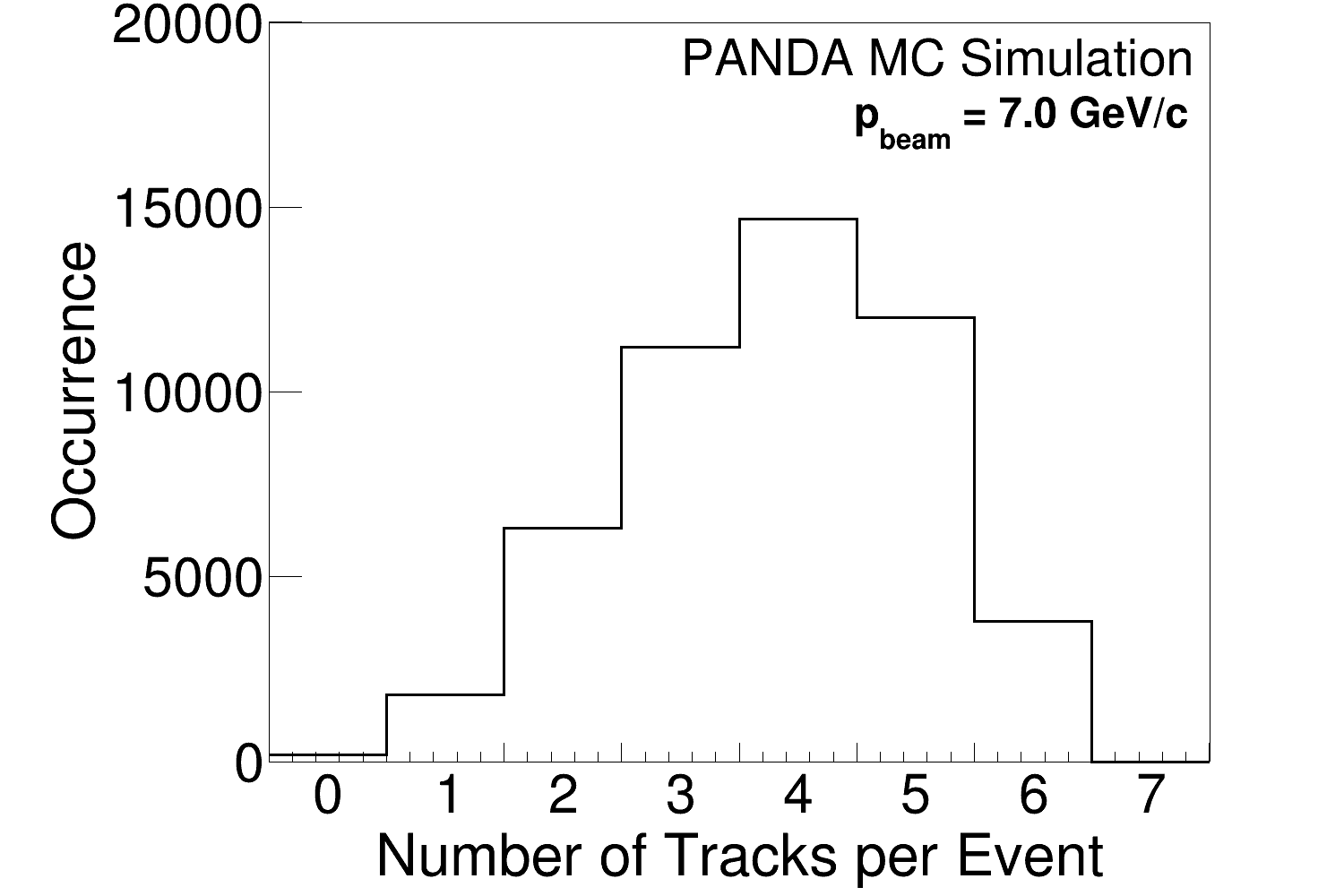}
\includegraphics[width=0.32\textwidth]{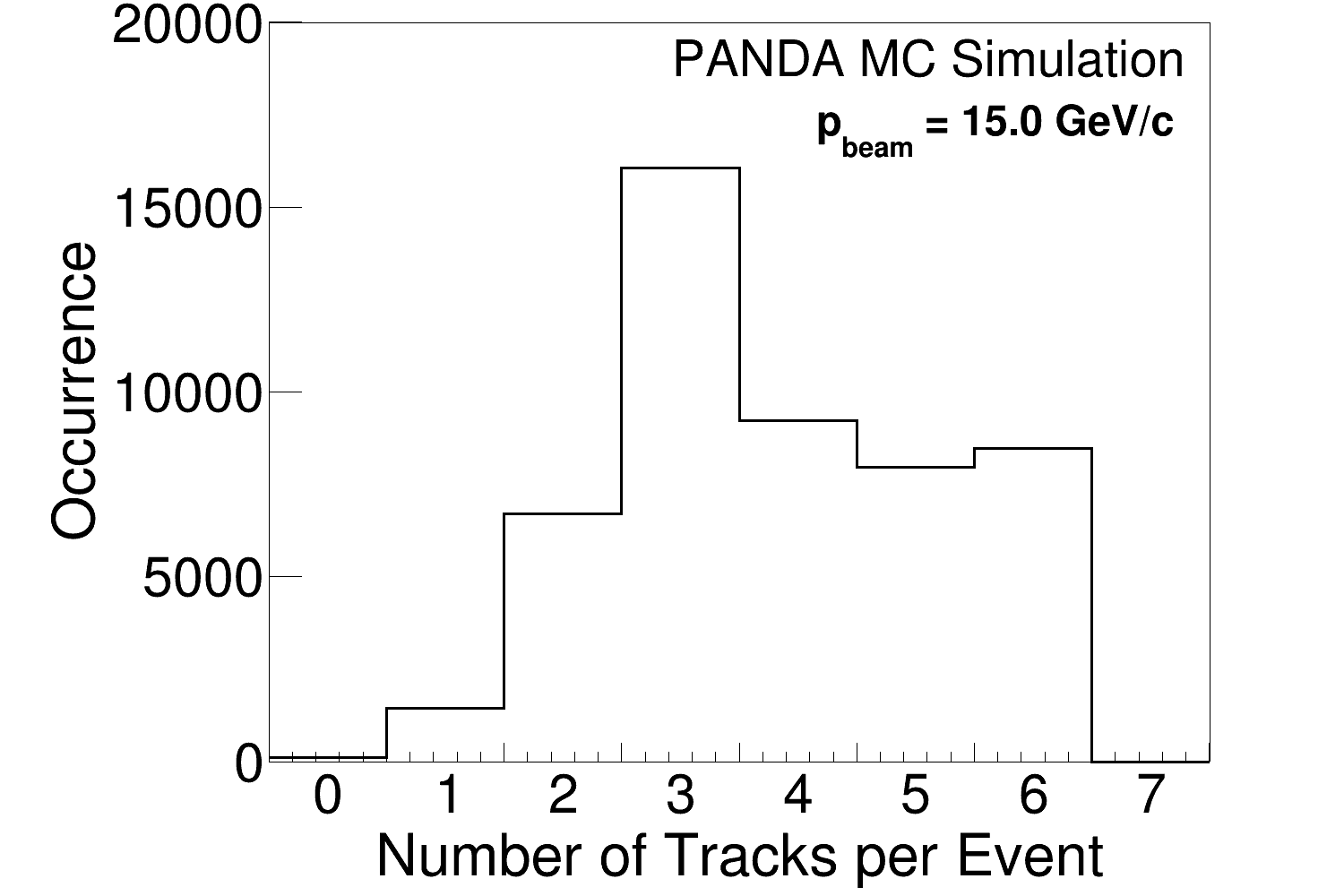}
\end{center}
\caption{The number of reconstructable TS tracks per event for 50,000 events at beam momenta 5.0 GeV/\textit{c} (left panel), 7.0 GeV/\textit{c} (middle panel) and 15.0 GeV/\textit{c} (right panel).}
\label{fig:RecoTracksOmega}
\end{figure}

The number of reconstructable tracks per event in the TS are shown in Fig. \ref{fig:RecoTracksOmega}. At 5.0 GeV/\textit{c}, about two TS tracks are reconstructed per event while at larger beam momenta, most event contain more three or more reconstructable TS tracks.

\begin{table}[h!]
\caption{The fraction of reconstructable final state tracks emitted within the TS from the $\bar{p}p \to \bar{\Omega}^+\Omega^-$ reaction. The normalization is the number of generated particles.}
\begin{center}
\begin{tabular}{| c | c | c | c |}
\hline
Particle & 5.0 GeV/\textit{c} &  7.0 GeV/\textit{c} & 15.0 GeV/\textit{c} \\
\hline
$p$ & 76\% & 67\% & 64\% \\
$\bar{p}$ & 76\% & 66\% & 64\% \\ 
$K^-$ & 79\% & 59\% & 64\% \\
$K^+$ & 80\% & 61\% & 66\%\\ 
$\pi^-$ & 80\% & 63\% & 61\%\\
$\pi^+$ & 81\% & 63\% & 62\% \\
\hline
\end{tabular}
\end{center}
\label{tab:ResultsTracksTS}
\end{table}

\subsection{Tracking detector hits}
\label{sec:trackhits}

\noindent The tracking of $\Lambda$ decay products originating from inside the STT rely on STT-based algorithms \textit{e.g.} those described in Ref. \cite{PzFinder}. The same is true for particles originating from the area between the MVD and the first GEM station, provided they are emitted in the direction of the STT. The transverse momentum components $p_x$ and $p_y$ are obtained from a preliminary track fit, which requires at least three hits. However, requiring at least four hits in the STT gives a more accurate fit. From Table \ref{tab:HitsInSTT}, we conclude that at 7.0 and 15.0 GeV/\textit{c}, most particles emitted within the TS acceptance (first row) give rise to 4 STT hits or more (second row).

\begin{figure}[h!]
\begin{center}
\includegraphics[width=.32\textwidth]{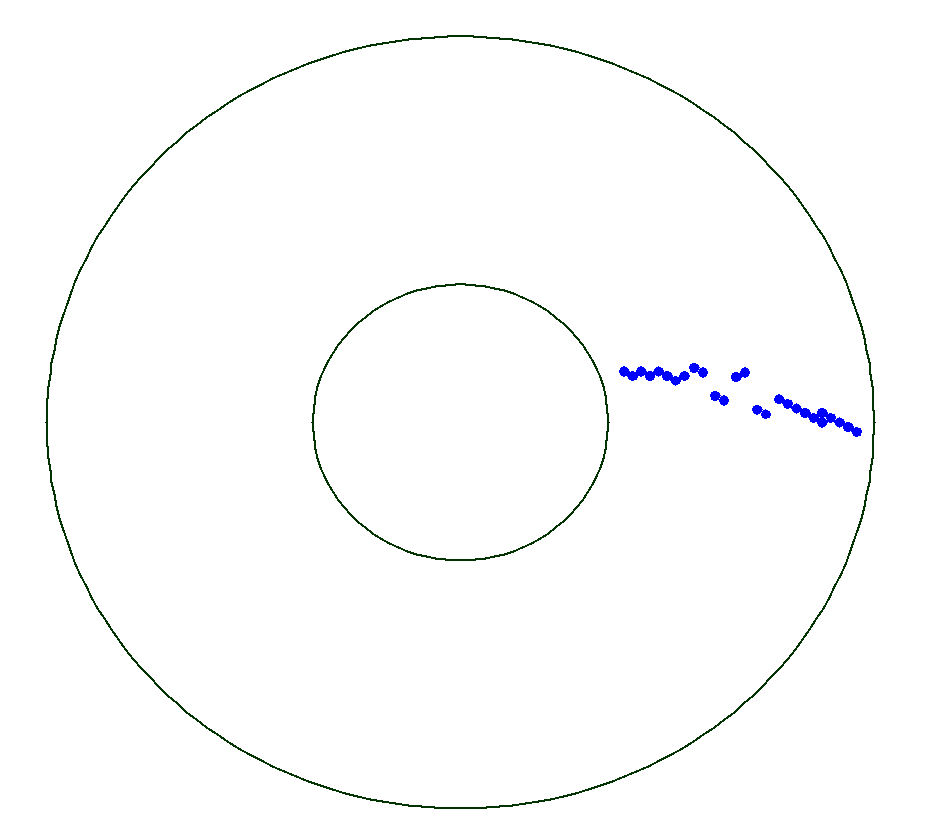}
\includegraphics[width=.32\textwidth]{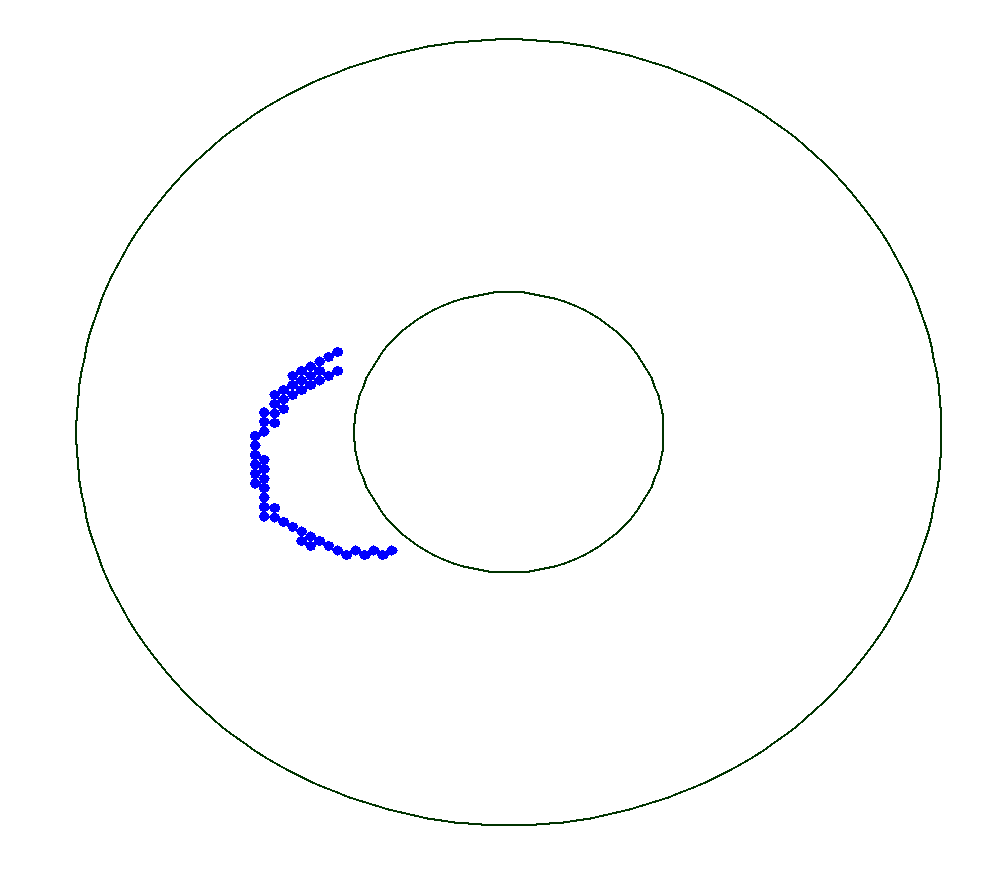}
\includegraphics[width=.32\textwidth]{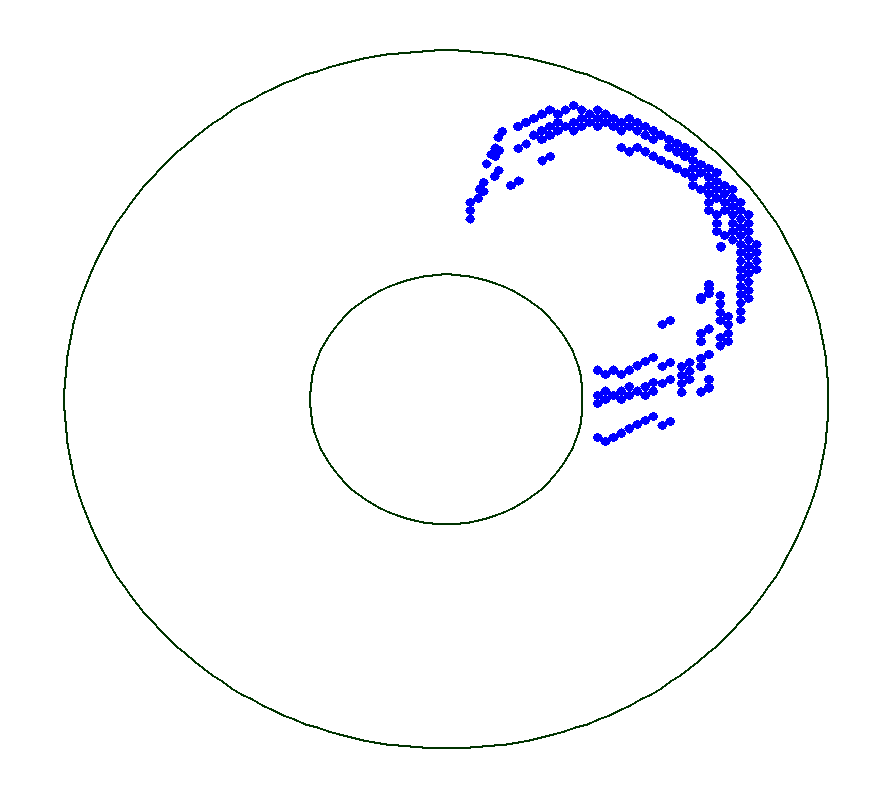}
\end{center}
\caption{Projection of the STT hits onto the xy-plane for the case when the particle traverses the full STT, leaving a hit in all 27 layers (left) and the case when the particle is trapped by the solenoid field in a spiral trajectory due to its small transverse momentum (right). Missing hits are due the target beam pipe.}
\label{fig:EventDisplaySTT}
\end{figure}

\begin{figure}[h!]
\begin{center}
\includegraphics[width=.32\textwidth]{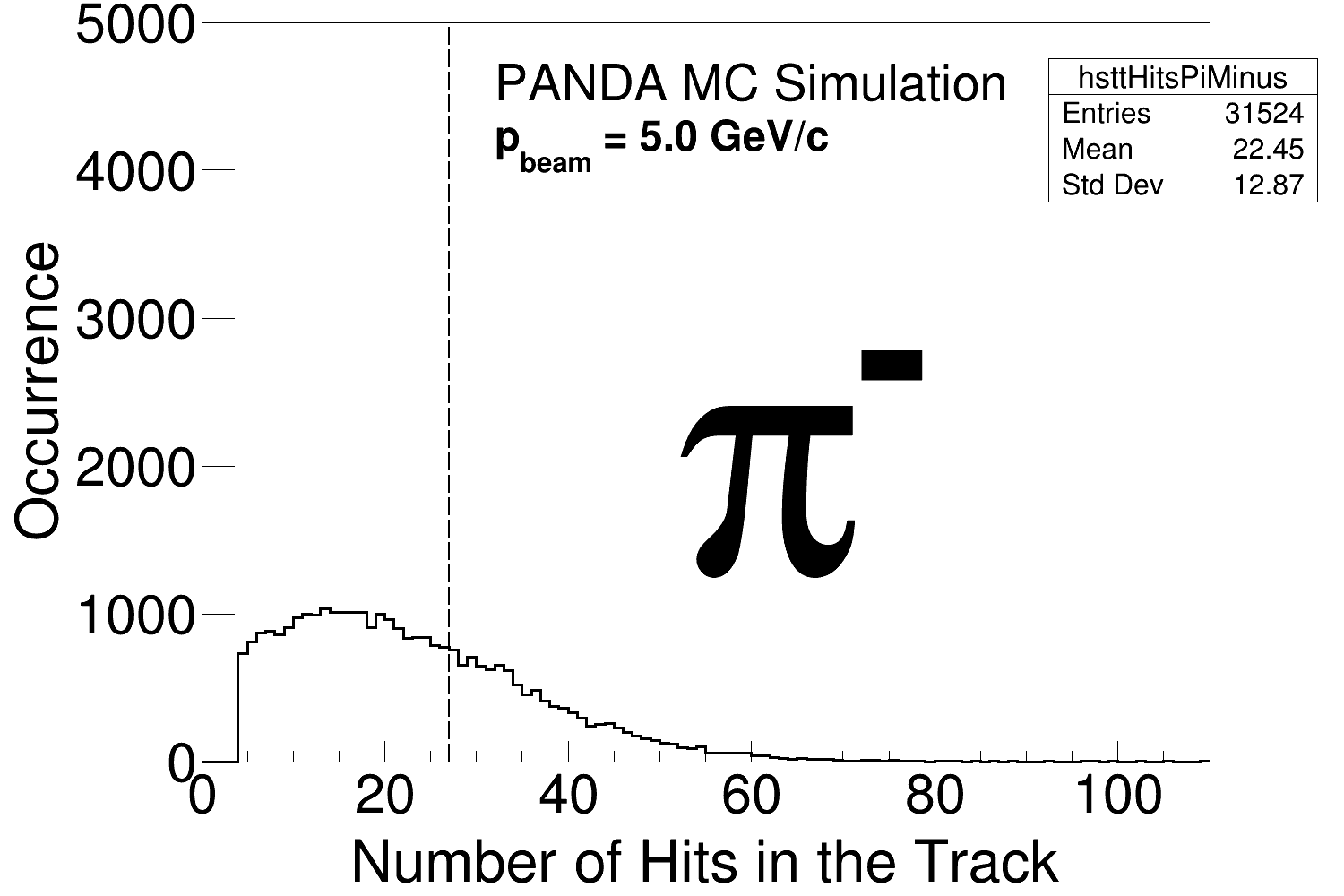}
\includegraphics[width=.32\textwidth]{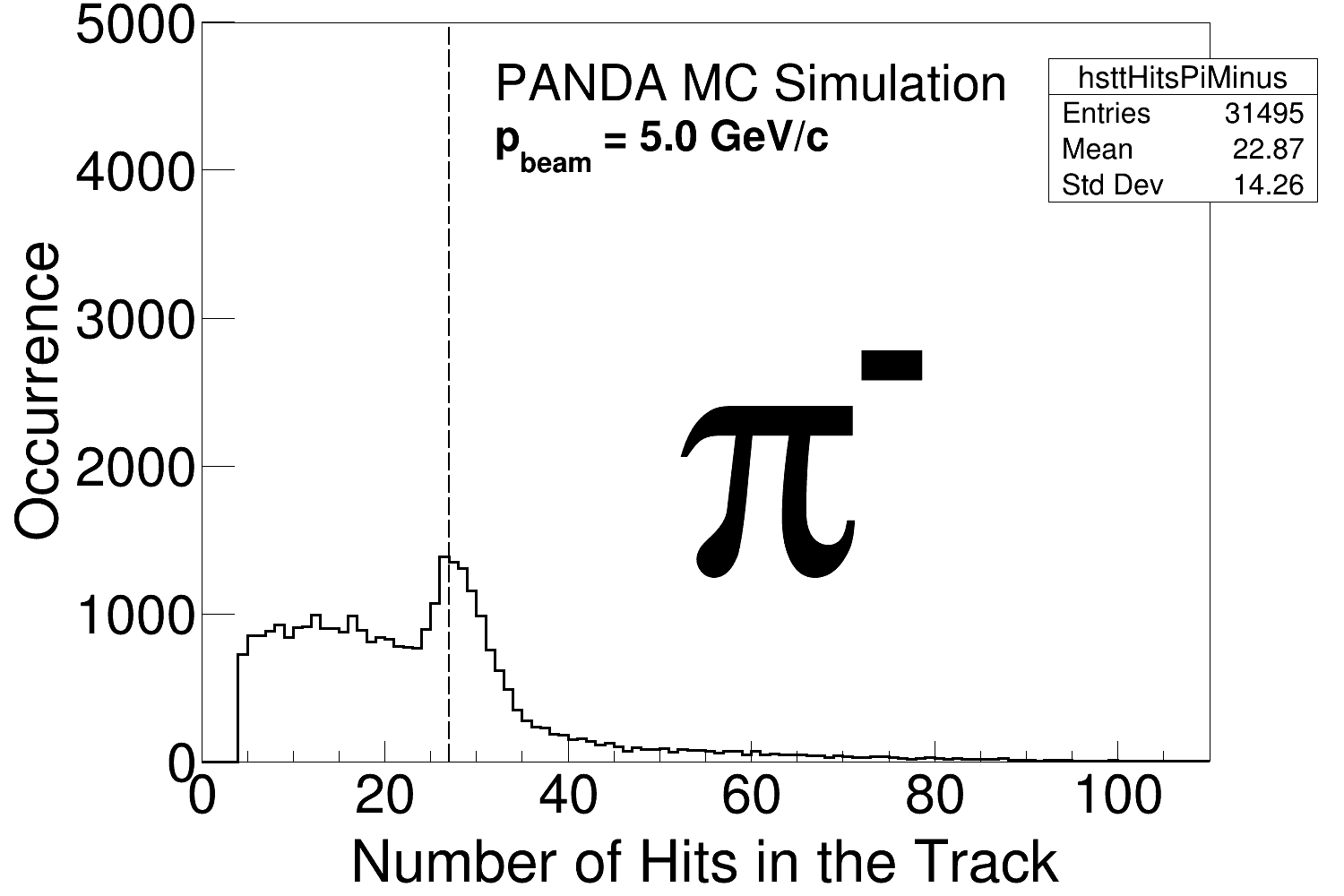}
\includegraphics[width=.32\textwidth]{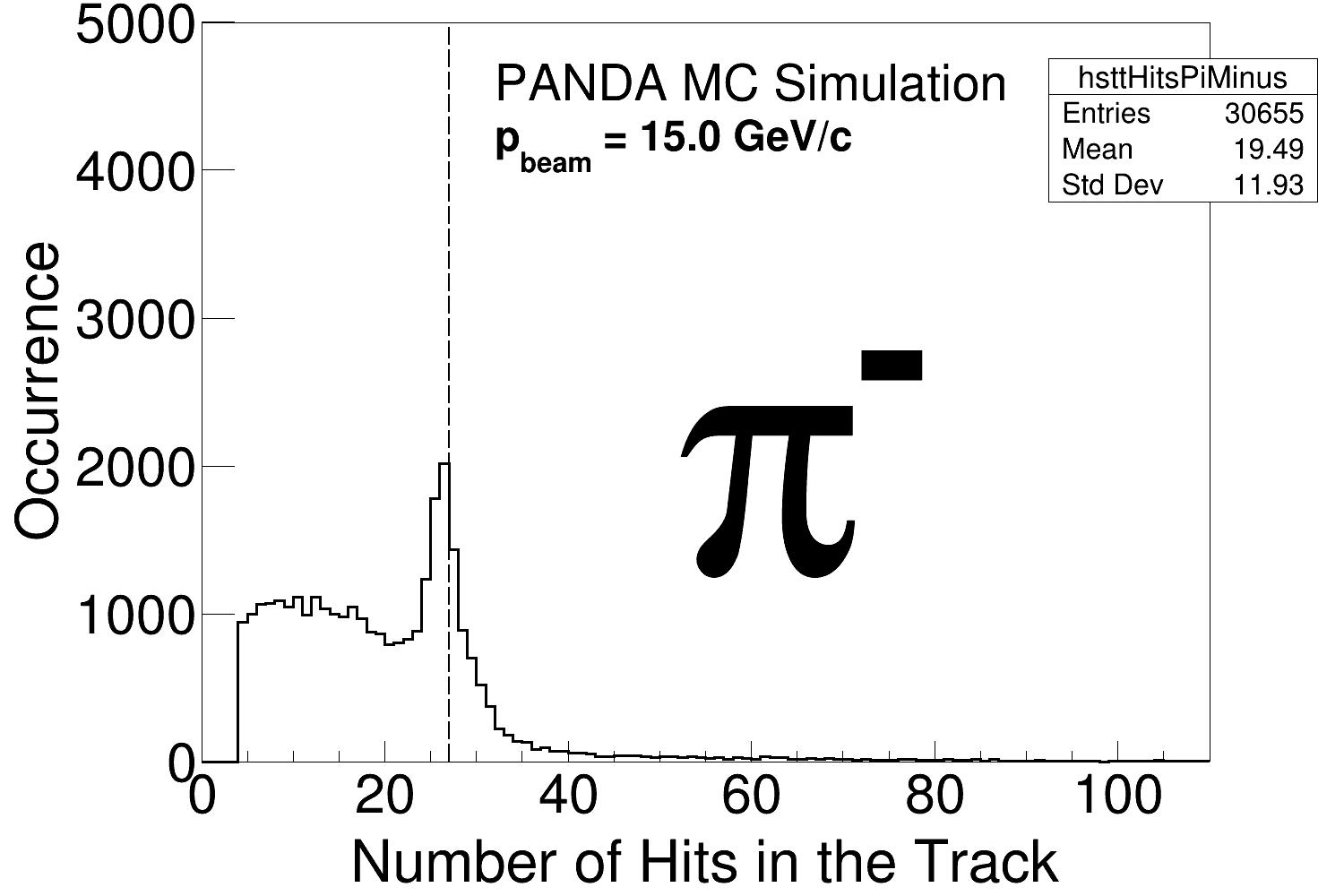}
\end{center}
\caption{Distribution of the number of STT hits from $\pi^-$ from the $\bar{\Omega}^+\Omega^-, \Omega^- \to \Lambda K^-, \Lambda \to p \pi^-$ process at 5.0, 7.0 and 15.0 GeV/$c$. The number of straw tube layers in the STT is 27 and marked with a vertical line.}
\label{fig:STTPions}
\end{figure}

The ideal STT hit pattern to fit is when a particle traverses the full STT in radial direction and leaves a hit in each of the 27 layers. This is illustrated in the left panel of Fig. \ref{fig:EventDisplaySTT}. Particles with small transverse momenta, which we know from Fig. \ref{fig:momentumAngle}- \ref{fig:2DMomentum} are primarily pions, could however get trapped in the solenoid field where they lose energy ($\approx$ 10 keV/cm) through interactions with the gas and tube walls in the STT. As a consequence, these pions will move in spiral trajectories and hence give rise to a very large number of hits. This is illustrated in the middle and right panels of Fig. \ref{fig:EventDisplaySTT}. Pions making a full turn or more, hence leaving $\geq$ 50 hits in the STT, are difficult to reconstruct with common algorithms. In Fig. \ref{fig:STTPions}, the hit pattern for pions is shown. At 7.0 and 15.0 GeV/\textit{c}, the STT hit distribution features a clear peak around 27 hits which corresponds to pions traversing the full STT. However, at 5.0 GeV/\textit{c}, no such peak is seen which means that many of these pions are either stopped or trapped within the STT. However, we see in Table \ref{tab:HitsInSTT} that the number of events with pions leaving more than 50 hits is lower than 1.1$\%$. 

\begin{figure}[h!]
\begin{center}
\includegraphics[width=.32\textwidth]{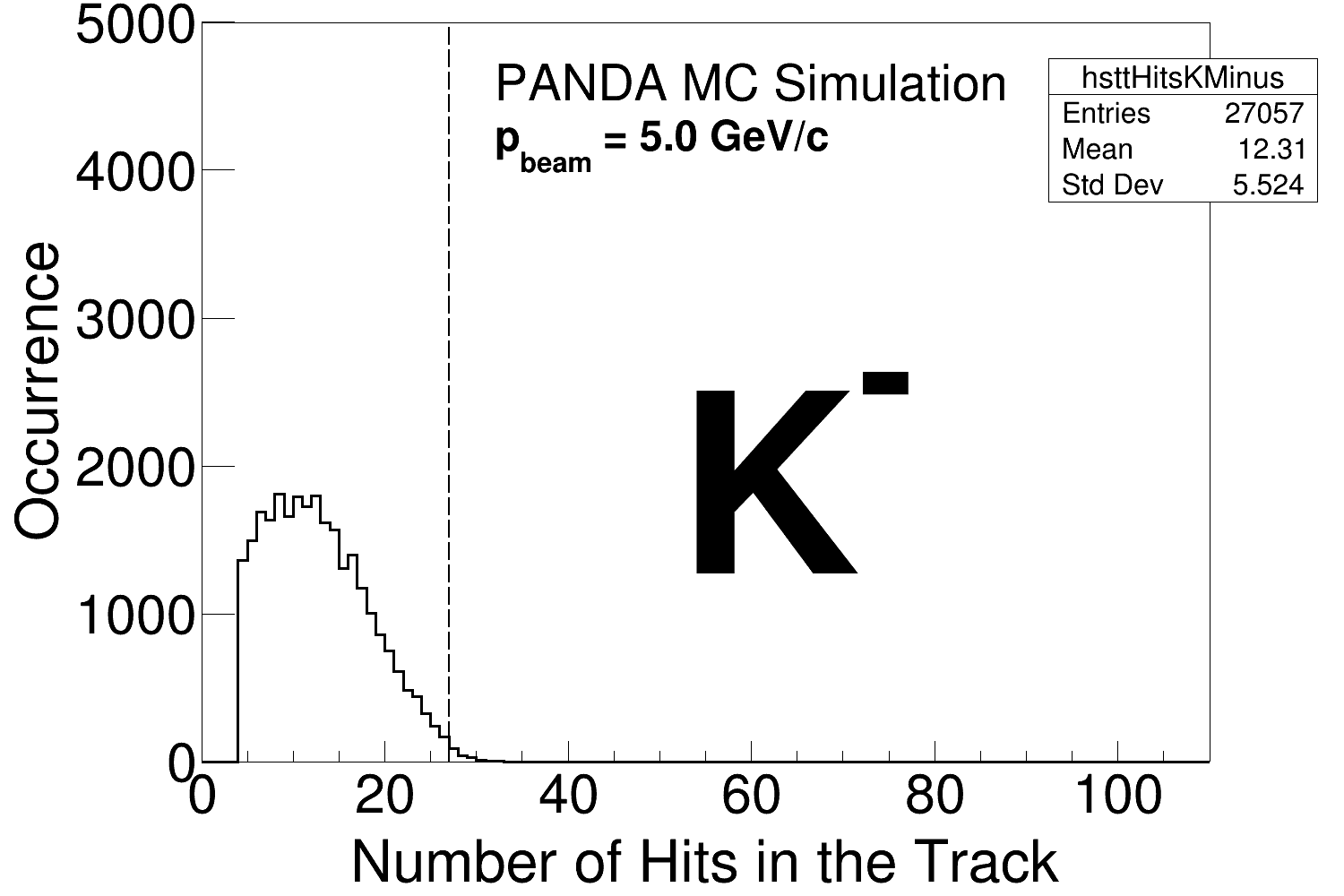}
\includegraphics[width=.32\textwidth]{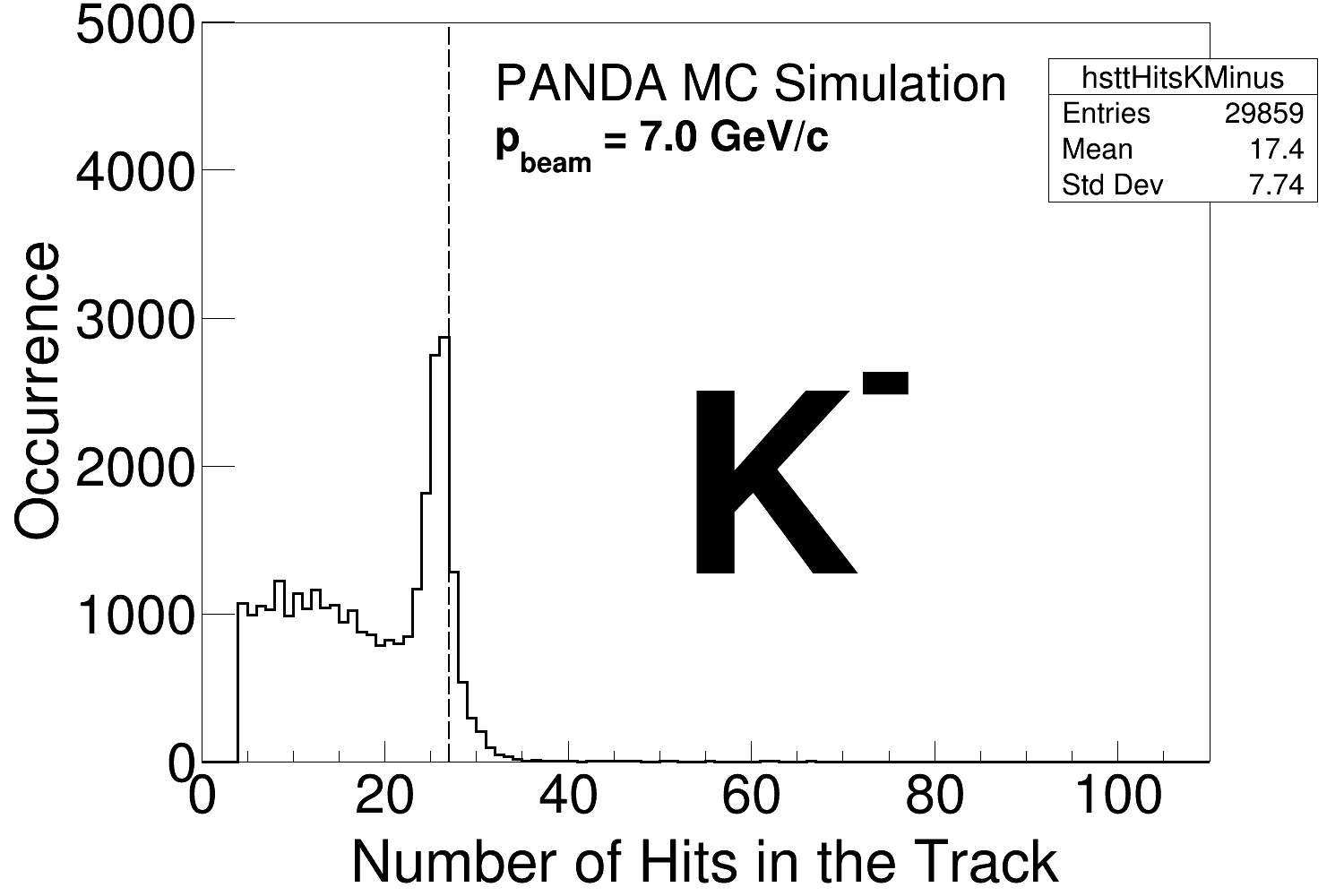}
\includegraphics[width=.32\textwidth]{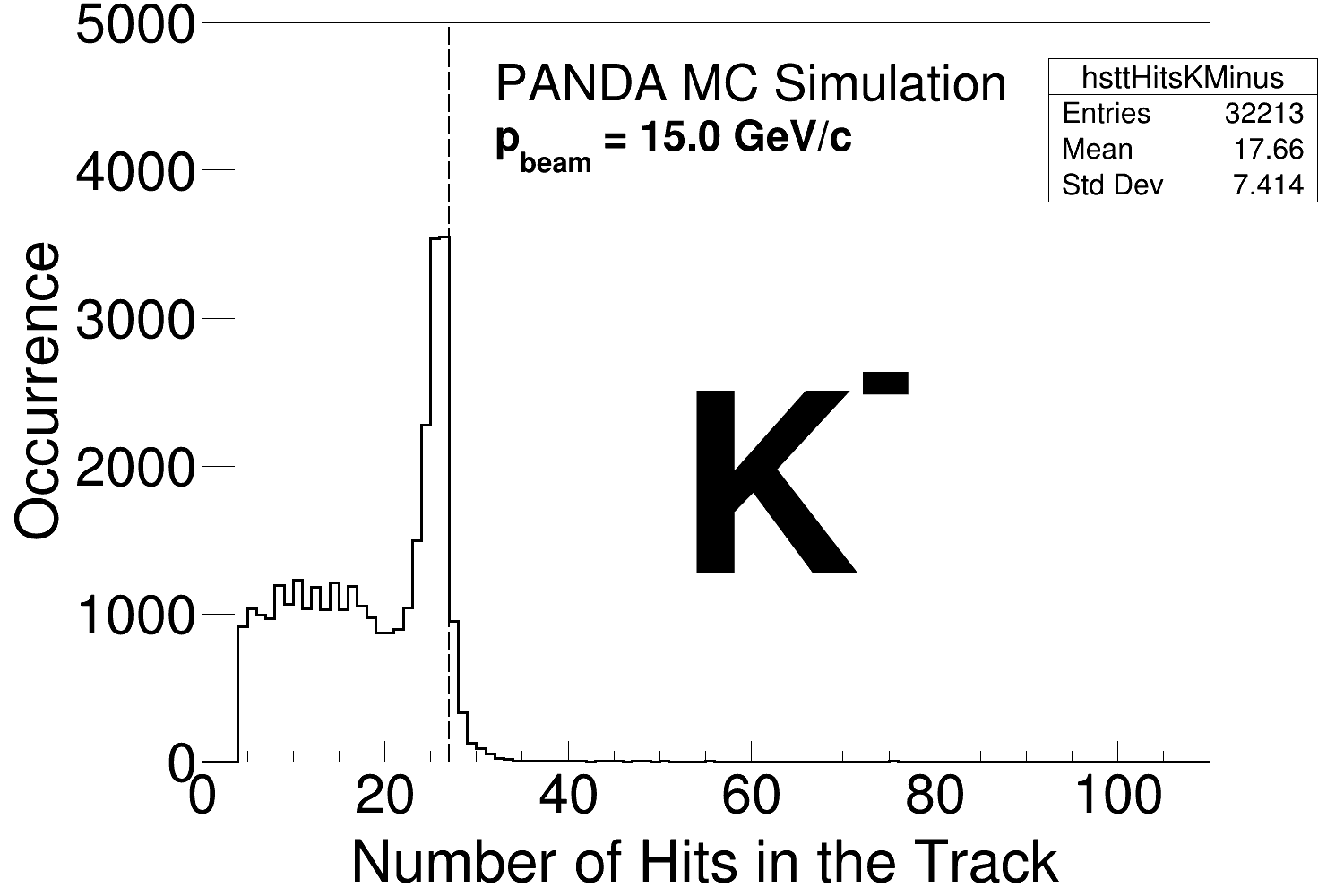}
\end{center}
\caption{Distribution of the number of STT hits from $K^-$ from the $\bar{\Omega}^+\Omega^-, \Omega^- \to \Lambda K^-$ process at 5.0, 7.0 and 15.0 GeV/$c$. The number of straw tube layers in the STT is 27 and marked with a vertical line.}
\label{fig:STTKaons}
\end{figure}

In Fig. \ref{fig:STTKaons}, the hit pattern from the kaons is shown. The clearly discernible peaks around 27 hits at the 7.0 GeV/\textit{c} and 15.0 GeV/\textit{c} show that many kaons traverse the full STT detector at these beam momenta.

\begin{table}[h!]
\caption{The fraction of reconstructable tracks from the $\bar{p}p \to \bar{\Omega}^+\Omega^-$ reaction with different number of hits in the STT.}
\begin{center}
\begin{tabular}{| l | c | c | c | }
\hline
STT hits / track & 5.0 GeV/\textit{c} & 7.0 GeV/\textit{c} & 15.0 GeV/\textit{c} \\
\hline
$\geq$ 4 & 43$\%$ & 63$\%$ & 64$\%$ \\
$\geq$ 50 & 0.7$\%$ & 1.1\% & 0.6$\%$ \\
\hline
\end{tabular}
\end{center}
\label{tab:HitsInSTT}
\end{table}

\subsection{Detector hits for time information}
\label{sec:timehits}

\noindent The number of events containing at least one signal from either the MVD, the GEM or the BarrelToF is shown in Table \ref{tab:ResultsEventsOmega}. In almost all events, there is a signal in either the MVD or the GEM, or both. Furthermore, many events also contain at least one hit in the dedicated timing detector, \textit{i.e.} the BarrelToF. We conclude that a $t_0$ time-stamp can be obtained from either the MVD, the GEM or the BarrelToF in practically all $\bar{p}p \to \bar{\Omega}^+\Omega^-$ events. 

\begin{table}[h!]
\caption{\textbf{Events with hits:} The fraction of all generated events from the $\bar{p}p \to \bar{\Omega}^+\Omega^-$ reaction with hits in different fast subdetectors.}
\begin{center}
\begin{tabular}{ | l | c | c | c | }
\hline
Detector & 5.0 GeV/\textit{c} & 7.0 GeV/\textit{c} & 15.0 GeV/\textit{c} \\
\hline
MVD & 99.5\% & 99.6\% & 97.5\%\\
GEM & 99.9\% & 99.8\% & 97.8\% \\
BarrelToF & 0.6\% & 40.0\% & 57.3\% \\
\hline
\end{tabular}
\end{center}
\label{tab:ResultsEventsOmega}
\end{table}

Table \ref{tab:ResultsTracksOmega} displays the fraction of the reconstructable tracks in the TS that have associated hits in fast BarrelToF, MVD or GEM detectors. This is crucial at higher luminosities, since it will then be necessary to associate each track segment with a reference time. Indeed, in most events, the tracks include hits from at least one of the fast detectors.

\begin{table}[h!]
\caption{\textbf{Tracks with hits:} The fraction of tracks within the TS acceptance with hits in fast detectors. The last row shows the fraction of the tracks within the TS acceptance.}
\begin{center}
\resizebox{\textwidth}{!}{
\begin{tabular}{ | l | c | c | c |}
\hline
Detector & 5.0 GeV/\textit{c} &  7.0 GeV/\textit{c} & 15.0 GeV/\textit{c} \\
\hline
MVD & 86\% & 74\% &  59\%\\
GEM & 87\% & 63\% & 51\%\\
MVD or GEM & 99\% & 93\%  & 84\% \\
BarrelToF & 0.2\% & 11\% & 20\%\\ 
BarrelToF or MVD & 86\% & 76\% & 64\%\\
BarrelToF or MVD or GEM & 99\% & 94\% & 89\% \\
\hline
Fraction of all generated tracks (in TS) & 43\% & 63\%  & 64$\%$ \\
\hline
\end{tabular}}
\end{center}
\label{tab:ResultsTracksOmega}
\end{table}

\subsection{Track efficiency}
\label{sec:acc}

\begin{figure}[h!]
\begin{center}
\includegraphics[width=.90\textwidth]{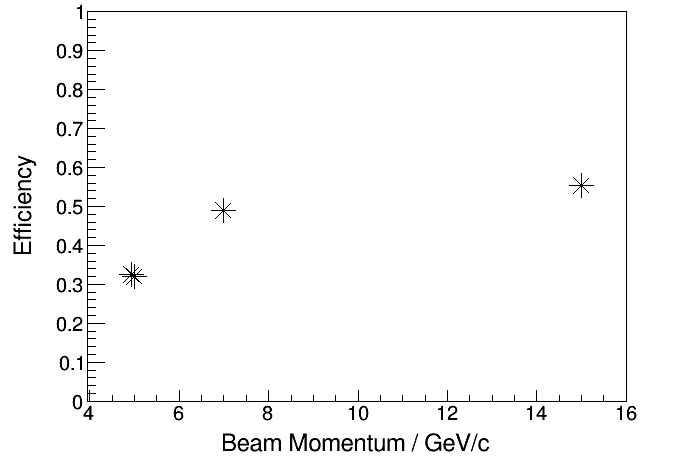}
\end{center}
\caption{Event reconstruction efficiency of the $\bar{p}p \to \bar{\Omega}^+\Omega^-$ reaction and its subsequent decays.}
\label{fig:OmegaAcceptance}
\end{figure}

\noindent In this first feasibility study of the $\bar{p}p \to \bar{\Omega}^+\Omega^-$ reaction with PANDA, the track efficiency in the TS and the FS was determined. The criterion for track reconstructability in the FS is at least six hits in the FTS planes. In addition to measure the number of final state particles (pions, kaons and protons/antiprotons), we also combine pions and protons (antiproton) into $\Lambda$ ($\bar{\Lambda}$) candidates as well as kaons and $\Lambda$ ($\bar{\Lambda}$) into $\Omega^-$ ($\bar{\Omega}^+$) candidates. Finally, we calculate the number of event where we have a pair of an $\Omega^-$ and an $\bar{\Omega}^+$ candidate. The results are presented in Table \ref{tab:ResultsTracksOmegaParticles}. The number of generated events, 50,000, are used for normalization. The final $\Omega^-\bar{\Omega}^+$ efficiency is presented in Fig.~\ref{fig:OmegaAcceptance} and Table \ref{tab:ResultsTracksOmegaParticles}. It is found to be larger than 30\% at all studied beam momenta, even very close to the kinematic threshold. The main variation with respect to the beam momentum is associated with the kaon reconstruction. 

\begin{table}[h!]
\caption{The track acceptance in the full PANDA detector from the $\bar{p}p \to \bar{\Omega}^+\Omega^-$ reaction and its subsequent decays.}
\begin{center}
\begin{tabular}{| c | c | c | c | c |}
\hline
Particle & 4.94 GeV/\textit{c} & 5.0 GeV/\textit{c} &  7.0 GeV/\textit{c} & 15.0 GeV/\textit{c} \\
\hline
$p$ & 87$\%$ & 87$\%$ & 92$\%$ & 88$\%$ \\
$\bar{p}$ & 83$\%$ & 84$\%$ & 91$\%$ & 87$\%$ \\ 
$K^-$ & 80$\%$ & 80$\%$ & 87$\%$ & 90$\%$ \\
$K^+$ & 82$\%$ & 82$\%$ & 88$\%$ & 91$\%$ \\ 
$\pi^-$ & 82$\%$ & 82$\%$ & 87$\%$ & 86$\%$ \\
$\pi^+$ & 83$\%$ & 83$\%$ & 86$\%$ & 87$\%$ \\
\hline
$\Lambda$ & 72$\%$ & 72$\%$ & 80$\%$ & 81$\%$\\
$\bar{\Lambda}$ & 69$\%$ & 69$\%$ & 79$\%$ & 80$\%$ \\
\hline
$\Omega^-$ & 57$\%$ & 57$\%$ & 70$\%$ & 73$\%$ \\
$\bar{\Omega}^+$ & 57$\%$ & 56$\%$ & 70$\%$ & 74$\%$ \\
\hline
$\Omega^-\bar{\Omega}^+$ & 33$\%$ & 32$\%$ & 49$\%$ & 55$\%$ \\
\hline
\end{tabular}
\end{center}
\label{tab:ResultsTracksOmegaParticles}
\end{table}

\section{Summary and Outlook}
\label{sec:sumout}

\noindent We have studied the reaction $\bar{p}p \to \bar{\Omega}^+\Omega^-$ and its subsequent decays in the PANDA detector, with focus on the TS. By studying hit patterns in tracking detectors and hits in fast detectors, we achieve a detailed understanding of the reconstruction feasibility of this reaction. This is crucial when developing track reconstruction and when outlining a strategy for event building and triggering. 

From the studied quantities we can formulate criteria in terms of particle kinematics, decay vertex positions, hits from detectors providing tracking and time information, and track efficiency. Our results demonstrate the importance of the MVD and GEM detectors for time-stamps; these detectors are hit in at least 97.5$\%$ of the events. The STT detector is crucial in the tracking of hyperon decay products since more than 43$\%$ of all tracks contain at least four STT hits.

The track efficiency of the PANDA detector (TS + FS) was determined from the acceptance in simulations and was found to increase from about 30\% close to threshold to about 50\% at larger beam momenta. This indicates a good feasibility for studies of the $\bar{p}p \to \bar{\Omega}^+\Omega^-$. The final efficiency will however be lower, after taking finite trigger- detector and reconstruction efficiencies into account.

Finally, our study provides guidance in how to exploit the hyperon detector signatures in the software trigger and event building scheme:

\begin{itemize}

\item Kaons from the $\Omega$ decays can be reconstructed using a primary track finder, \textit{i.e.} an algorithm that uses the constraint that the track originates from the interaction point, combining information from the MVD and the STT.
\item The $\Lambda$ hyperons must be reconstructed in the STT or the FTS utilizing a secondary track finder.
\item At least three reconstructed TS tracks are required in each event.
\item Up to four displaced decay vertices in the MVD, the STT or in the FS are required in each event.
\item The $t_0$ of TS tracks are obtained from the MVD, GEM or BarrelToF detectors.
\end{itemize}

\noindent These studies demonstrates the potential of a program of pioneering studies of the production and decay of $\Omega$ hyperons, and demonstrates the versatility and good performance of the PANDA detector.

\section{Acknowledgements}

\noindent We acknowledge financial support from
  the Bhabha Atomic Research Centre (BARC) and the Indian Institute of Technology Bombay, India;
  the Bundesminis\-terium f\"ur Bildung und Forschung (BMBF), Germany;
  the Carl-Zeiss-Stiftung 21-0563-2.8/122/1 and 21-0563-2.8/131/1, Mainz, Germany;
  the Center for Advanced Radiation Technology (KVI-CART), Groningen, Netherlands;
  the CNRS/IN2P3 and the Universit\'{e} Paris-Sud, France;
  the CU (Czech Republic): MSMT LM2018112, OP VVV CZ.02.1.01/0.0/0.0/18\_046/0016066;
  the Deutsche Forschungsgemeinschaft (DFG), Germany;
  the Deutscher Akademischer Austauschdienst (DAAD), Germany;
  the European Union's Horizon 2020 research and innovation programme under grant agreement No 824093;
  the Forschungszentrum J\"ulich, Germany;
  the Gesellschaft f\"ur Schwerionenforschung GmbH (GSI), Darmstadt, Germany;
  the Helmholtz-Gemeinschaft Deutscher Forschungszentren (HGF), Germany;
  the INTAS, European Commission funding;
  the Institute of High Energy Physics (IHEP) and the Chinese Academy of Sciences, Beijing, China;
  the Istituto Nazionale di Fisica Nucleare (INFN), Italy;
  the Ministerio de Educaci\'on y Ciencia (MEC) under grant FPA2006-12120-C03-02, Spain;
  the Polish Ministry of Science and Higher Education (MNiSW) grant No. 2593/7, PR UE/2012/2, and the National Science Centre (NCN) DEC-2013/09/N/ST2/02180, Poland;
  the Schweizerischer Natio\-nalfonds zur F\"orderung der Wissenschaftlichen Forschung (SNF), Switzerland;
  the Science and Technology Facilities Council (STFC), British funding agency, Great Britain;
  the Scientific and Technological Research Council of Turkey (TUBITAK) under the Grant No. 119F094, Turkey;
  the Stefan Meyer Institut f\"ur Subatomare Physik and the \"Osterreichische Akademie der Wissenschaften, Wien, Austria;
  the Swedish Research Council and the Knut and Alice Wallenberg Foundation, Sweden;
  the U.S. Department of Energy, Office of Science, Office of Nuclear Physics;
  \\
  \newline 
  The results published in this article have been obtained in collaborative
efforts before 24.02.2022 and are thus signed by the full collaboration
including scientists from Russian and Belarus institutions.

 \bibliographystyle{elsarticle-num}

\end{document}